\documentclass[aps,prb,twocolumn,floatfix,,
superscriptaddress]{revtex4-2}
\usepackage{amssymb}
\usepackage{amsmath}
\usepackage{graphicx}
\usepackage{textcomp}
\usepackage{color}

\begin{document}

\title{Magnetostriction of metals with small Fermi surface pockets: Case of the topologically trivial semimetal LuAs}
		
\author{Yu.\,V.\,Sharlai}
\affiliation{B. Verkin Institute for Low Temperature Physics and Engineering of the National Academy of Sciences of Ukraine, Kharkiv 61103, Ukraine}
\affiliation{Institute of Low Temperature and Structure Research, Polish Academy of  Sciences, 50-422 Wroc{\l}aw, Poland}
\author{L.\,Bochenek}
\affiliation{Institute of Low Temperature and Structure Research, Polish Academy of  Sciences, 50-422 Wroc{\l}aw, Poland}
\author{J.\,Juraszek}
\affiliation{Institute of Low Temperature and Structure Research, Polish Academy of  Sciences, 50-422 Wroc{\l}aw, Poland}
\author{T.\,Cichorek}
\affiliation{Institute of Low Temperature and Structure Research, Polish Academy of  Sciences, 50-422 Wroc{\l}aw, Poland}
\author{G.\,P.\,Mikitik}
\affiliation{B. Verkin Institute for Low Temperature Physics and Engineering of the National Academy of Sciences of Ukraine, Kharkiv 61103, Ukraine}

 \begin{abstract}
We develop a theory of the magnetostriction for metals with small charge-carrier pockets of their Fermi surfaces. As an example, we consider LuAs that has a cubic crystal structure. The theory quite well describes the known experimental data on the magnetostriction of this metal. The obtained results also clearly demonstrate that the dilatometry can be used for detecting tiny Fermi-surface pockets that are not discerned by traditional methods based on investigations of the Shubnikov--de Haas and de Haas--van Alphen effects.
 \end{abstract}

\maketitle

\section{Introduction}

Magnetostriction (MS) of nonmagnetic conductive materials is directly related to changes in the density of charge carriers in a magnetic field, and this thermodynamic quantity was studied theoretically and experimentally in numerous works, see, e.g., Refs.~ \cite{Keyes,fawcett,Sh,Mi,Kuchler1,m-sh15,lif,juraszek}. Recently, it was shown that measurements of field-induced length change is an effective probe for detecting Weyl  electrons in topological semimetals because even in moderate magnetic fields, which are too weak to confine large groups of massive quasi-particles at their zeroth Landau levels, the MS contains a linear-in-field term that identifies the presence of relativistic fermions \cite{cichorek}. Another profound advantage of the magnetostriction lies in its increase with decreasing the charge-carrier density, and hence making it well suited  for investigation of small charge-carrier groups also in ordinary (nontopological) materials. In this paper, on the example of diamagnetic semimetal LuAs with a trivial electronic band structure \cite{juraszek}, we demonstrate how the dilatometry can be used to discern and analyze small charge-carrier groups, which are not detected by both quantum oscillations and the angle-resolved photoemission spectroscopy \cite{juraszek}.

LuAs belongs to the family of nonmagnetic monopnictides which crystallize in a rock-salt structure (space group \textit{Fm-3m}). The rediscovery of extremely large magnetoresistance (XMR) in nonmagnetic monopnictides has drawn attention to these binary compounds with simple Fermi surfaces (FSs) \cite{Kitazawa,Tafti1,Tafti2}, as ideal model systems to investigate the field-induced properties of semimetals. Different scenarios for an origin of the XMR have been proposed, including the band inversion and the orbital texture of the electron pockets (see the inset of Fig.~\ref{fig1}) \cite{Nummy}. These scenarios, however, do not apply to the topologically trivial compound LuAs which displays the nonsaturating XMR with a nonquadratic magnetic-field dependence gained even up to nearly $60$ T. Furthermore, this material exhibits the very large magnetostriction, indicative of a field-dependent variation of electron and hole concentrations (see Fig.~\ref{fig1}). Thus, previous findings on LuAs showed that certain high-field properties of topological semimetals may not necessarily reflect the presence of massless quasiparticles \cite{juraszek}.

The paper is structured as follows:
In the next section, we refer to the experimental results of Ref.~\cite{juraszek} on the magnetostriction of LuAs and emphasize  specific features of these findings. In Sec.~III, these  experimental data are analyzed, and we demonstrate how their specific features enable us to discover a tiny charge-carrier group which was not observed by other methods in LuAs. In Sec.~IV, we present new magnetostriction and magentoresistance results (obtained as described elsewhere \cite{juraszek}), which are in favor of the existence of the small group in LuAs, as well as we point out additional experiments that could extend possibility of the magnetostriction in analyzing the characteristics of small Fermi-surface pockets.
In Conclusions, the main results are summarized. A derivation of the necessary formulas for the magnetostriction is given in Appendices \ref{A}-\ref{C}. In Appendix \ref{D}, relations between the pressure dependences of the quantum-oscillation frequencies and  the constants of the deformation potential are presented.

\section{Experimental results} \label{II}

\begin{figure}[bp] 
 \centering  \vspace{+9 pt}
\includegraphics[scale=1.]{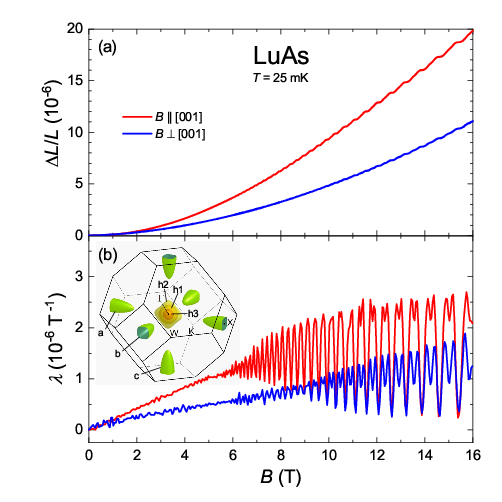}
\caption{\label{fig1} Top: Magnetostriction of LuAs measured along
the [001] axis for the magnetic field $B$ parallel to the [001] (red  line) and [100] (blue line) directions \cite{juraszek}. Bottom: The quantity $\lambda\equiv d(\Delta L/L)/dB$ as a function of $B$. Inset: The Brillouin zone of LuAs and its Fermi-surface pockets  \cite{juraszek}: $a$, $b$, $c$ mark electrons, $h1$, $h2$, $h3$ denote the hole charge carriers.
 } \end{figure}

As reported previously in Ref. \cite{juraszek}, the field-induced relative length change $\Delta L/L$  of the LuAs single crystal was measured in the longitudinal and transverse configurations when the magnetic field $B$ was parallel to the ${\bf z}=$[001] and ${\bf x}=$[100] axes, respectively  (Fig.~\ref{fig1}).
Since for a cubic-structure crystal, the magnetostriction along the $z$ axis at $B$\,$\parallel$\,$x$ coincides with the MS along the  $x$ axis  at $B$\,$\parallel$\,$z$, we shall discuss the $\Delta L/L$ along the $z$ and $x$ axes at fixed $B$\,$\parallel$\,$z$ in order to describe the experimental data obtained for the two direction of the magnetic field. These $\Delta L/L$ coincide with the $u_{zz}$ and $u_{xx}$ components of the strain tensor $u_{ik}$.

\begin{figure}[tbp] 
 \centering  \vspace{+9 pt}
\includegraphics[scale=0.99]{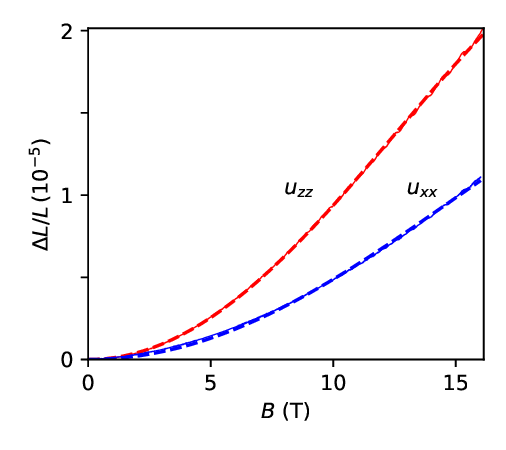}
\caption{\label{fig2} The magnetic-field dependences of magnetostriction of LuAs along the $z$ axis, $u_{zz}$, (red solid line)  and along the $x$, $u_{xx}$, (blue solid line) at $B\parallel z$, cf. Fig.~\ref{fig1}. The dashed lines show  the approximation of these dependences by Eq.~(\ref{1}) where $c^{z}_{2}\approx 1.047\times 10^{-7}$\ T$^{-2}$, $c^{z}_{4}\approx -1.102\times 10^{-10}$\ T$^{-4}$ for $u_{zz}$, and   $c^{x}_{2}\approx 5.28\times 10^{-8}$\ T$^{-2}$, $c^{x}_{4}\approx -4.05\times 10^{-11}$\ T$^{-4}$ for $u_{xx}$.
 } \end{figure}

The magnetostriction of LuAs is quite large, and it reaches  $2\times 10^{-5}$ in a magnetic field of $16$ T (Fig.~\ref{fig1}). This value exceeds  the magnetostriction of pure Bi \cite{Mi} ($1.5\times 10^{-5}$) and of TaAs \cite{cichorek} ($0.5\times 10^{-5}$)  at the same field.
This fact immediately indicates that small charge-carrier groups seem to play an important role in the elongation of LuAs \cite{cichorek}.
In the region $B\le 16$ T, the oscillating part of the $B$ dependence  of the magnetostriction is relatively small [and it is better discernable in the quantity $\lambda\equiv d(\Delta L/L)/dB$]. However, the most important thing is that its smooth part {\it substantively} deviates from the $B^2$ law. Indeed, as shown by the dashed lines in Fig.~\ref{fig2}, the elongations of the sample along the $z$ and $x$ axes are sufficiently well approximated by the function,
 \begin{eqnarray} \label{1}
 \frac{\Delta L}{L}\approx c_{2}B^2 +c_{4}B^4,
 \end{eqnarray}
where $c_{2}$, $c_{4}$ are the constants (Fig.~\ref{fig2}). We note that the deviation of the MS from the $B^2$ dependence amounts to about  $10\%$ at $10$ T. Thus, there are two unusual features of the magnetostriction of the cubic semimetal LuAs: i) there is an essential anisotropy of the magnetostriction (the elongations along the $z$ axis is approximately two times larger than the elongation along $x$) and ii) the smooth part of the MS noticeably deviates from the $B^2$ law. It is these features that will be analyzed below.

\section{Analysis of the experimental results}

\begin{figure}[tb] 
 \centering  \vspace{+9 pt}
\includegraphics[scale=1.]{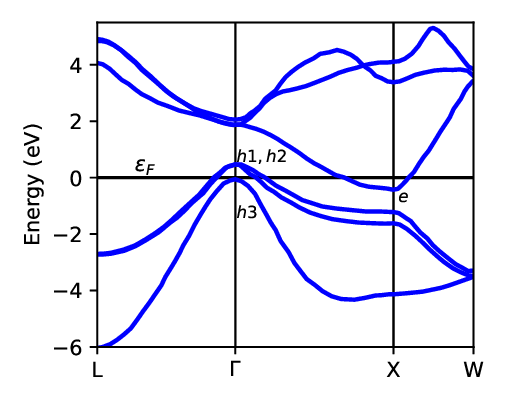}
\caption{\label{fig3} The electron-band structure of LuAs near its Fermi level $\varepsilon_F$. The structure is plotted on the basis of  the data of Ref.~ \cite{khalid}. The symbols $h1$, $h2$, $h3$, and $e$ mark the hole and electron bands, the charge carriers of which produce the Fermi surface shown in Fig.~\ref{fig1}.
 } \end{figure}   

\subsection{General considerations}

The electron-band-structure calculations \cite{juraszek,xie,khalid} show that there are relatively small electron and hole pockets of the Fermi surface in the Brillouin zone of LuAs  (Figs.~\ref{fig1} and \ref{fig3}). The centers of all the hole pockets $h1$, $h2$, $h3$ are at the center $\Gamma$ of its first Brillouin zone whereas the electron pockets $a$, $b$, $c$ are located at the $X$ points of the four-fold rotation axes $a$, $b$, $c$, which coincide with the $x$, $y$, $z$ axes of the crystal, respectively. Note that the existence of the tiny $h3$-hole pocket is still questionable, since this pocket has not yet been experimentally discovered \cite{juraszek,xie}.

As is known \cite{abr}, an elastic deformation shifts a band $\varepsilon({\bf p})$ in the Brillouin zone of the crystal, and this shift $\Delta\varepsilon({\bf p})$ generally depends on the quasimomentum ${\bf p}$ of electrons, $\Delta\varepsilon({\bf p})=\sum_{l,k}\lambda_{lk}({\bf p})u_{lk}$, where $u_{lk}$ is the strain tensor, and $\lambda_{lk}({\bf p})$ is the deformation potential of this band. However, for a small Fermi-surface pocket enclosing an extremal point ${\bf p}_0$ of $\varepsilon({\bf p})$, one may put  $\lambda_{lk}({\bf p})\approx \lambda_{lk}({\bf p}_0)$ since, in general, the deformation potential  $\lambda_{lk}({\bf p})$ changes significantly on the scale of the order of the size of the Brillouin zone. This constancy of $\lambda_{lk}$ is the so-called rigid-band approximation which is commonly used in describing the magnetostriction of crystals with small FS pockets \cite{Keyes,Mi,Kuchler1,m-sh15,lif,juraszek,cichorek}.
In this approximation, it is assumed that a deformation of a crystal shifts its electron-energy bands as whole without changing their shape. However, some caution is required when the band under study is separated from another band by a  sufficiently small gap $\Delta$. In this case, the effective electron masses of the band may be of the order of $\Delta/V^2$ where $V$ is a typical interband matrix element of the velocity operator ($V\sim 10^5-10^6$ m/s). If the small gap $\Delta$ substantially depends on a deformation, the rigid-band approximation fails even for a small pocket of this band. Indeed, in this case, the effective masses of the band, and hence its shape, noticeably change with the deformation \cite{c1}.
In LuAs, the holes bands are close to each other at the point $\Gamma$, and the electron band at the point X is not far away from the $h1$ and $h2$ bands (Fig.~\ref{fig3}). For this reason, although we will begin our analysis, implying the applicability of the rigid-band approximation to the small Fermi-surface pockets in LuAs, we shall not exclude a failure of this approximation.
Formulas for the magnetostriction of crystals with the cubic crystal structure are derived in Appendices \ref{A} - \ref{C} both in the rigid-band approximation and beyond it.

The applied magnetic field $B\parallel z$ singles out the $z$ axis of the crystal with the cubic structure, and therefore, generally speaking, $u_{zz}(B)\neq u_{xx}(B)=u_{yy}(B)$. In analyzing the experimental data, it is convenient to decompose the magnetostriction into the isotropic $(u_{zz}+2u_{xx})/3$ and anisotropic parts,
\begin{eqnarray*}
u_{zz}=\frac{u_{zz}+2u_{xx}}{3}+2\frac{u_{zz}-u_{xx}}{3},\\
u_{xx}=u_{yy}=\frac{u_{zz}+2u_{xx}}{3}-\frac{u_{zz}-u_{xx}}{3},
\end{eqnarray*}
and to start the analysis from the anisotropic part which is determines only by the electrons in LuAs. (As will be explained below, the hole pockets do not contribute to this part in the rigid-band approximation since their deformation potential has the cubic symmetry.)  This ``simplicity'' of the anisotropic part helps to reveal the problem of the theoretical description of the magnetostriction and indicates the way to solve it.

\subsection{Anisotropic part of the magnetostriction}\label{IIB}

\subsubsection{The anisotropic part within the rigid-band approximation.}

Within the rigid-band approximation, the hole charge carriers give an isotropic contribution to the magnetostriction, and its anisotropic part, $u_{zz}-u_{xx}$, is determined only by the electron Fermi-surface pockets [see Eqs.~(\ref{6b}) and (\ref{19b})],
\begin{eqnarray}\label{2}
u_{zz}-u_{xx}=(\Lambda_c^{\perp}-\Lambda_c^{\parallel})(\Delta n_a -\Delta n_c),
\end{eqnarray}
where $\Delta n_i\equiv n_i(B)-n_i(0)$ is the change in the charge-carrier density of the $i$th Fermi-surface pocket in the magnetic field (here $i=a$, $b$, $c$); the constants $\Lambda_c^{\parallel}$, $\Lambda_c^{\perp}$ are expressed in terms of the elastic moduli and the parameters of the deformation potential of LuAs, Eq.~(\ref{8b}), and we have taken into account in Eq.~(\ref{2}) that $\Delta n_a=\Delta n_b$.
Since the frequencies $F_i$ of the quantum oscillation of the electron pockets in LuAs are much larger than the  magnetic fields in experiments ($F_c\approx 280$ T, $F_a=F_b\approx 900$ T at $B\parallel z$ \cite{juraszek,xie}), the non-oscillating parts of $\Delta n_c$ and $\Delta n_a$ are proportional to $B^2$. If the electron in LuAs  are described by a parabolic dispersion, the $\Delta n_i$ are given by expressions (\ref{10b}),
\begin{eqnarray}\label{3}
\Delta n_i=\frac{3n_iB^2}{8F_i^2} \left(\delta_i^2- \frac{1}{12}\right),
\end{eqnarray}
where $n_i$ is the charge-carrier density in the $i$th pocket  at $B=0$, $\delta_i$ is the parameter characterizing the charge-carriers magnetic moment $\mu_i$\,=\,$\delta (e\hbar/m^*_{i})$, and  $m^*_{i}$ is their cyclotron mass. The moment $\mu_i$ consists of its spin and orbital parts, the latter being due to the spin-orbit interaction.

Let us estimate the value of $u_{zz}-u_{xx}$ for LuAs, using formulas (\ref{2}) and (\ref{3}).
With the frequencies $F_a$ and $F_c$, we can obtain the density $n_c$ of electrons in one of their pockets \cite{m-sh21a},
\begin{eqnarray}\label{4}
n_c=\frac{1}{3\sqrt{\pi}}\left( \frac{8F_{c}F_{a}^2}{\phi_0^3}\right)^{1/2} \approx 0.85\times 10^{20}  {\rm cm}^{-3},
\end{eqnarray}
where $\phi_0=hc/2e$ is the flux quantum.
The total electron density $n_e=3n_c\approx 2.55\times 10^{20}$   cm$^{-3}$ agrees with the value $n_e\approx 2.28\times 10^{20}$   cm$^{-3}$ calculated earlier in Ref.~\cite{juraszek} and is of the order of $n_e\approx 4.56\times 10^{20}$ cm$^{-3}$ estimated by Xie et al.\ \cite{xie}. At the point X of the Brillouin zone of LuAs, there are three close bands for which the energy gap without the spin-orbit coupling is comparable with the gap produced by the spin-orbit interaction \cite{khalid}. In this situation, the typical value of the parameter $\delta$ is of the order of unity, see Sec.~III of Ref. \cite{g1} in which a theory of the electron \textit{g} factor in metals was presented. Based on oscillatory MS \cite{juraszek}, it was found that the effective electron \textit{g} factor for the pocket $c$ is approximately equal to $6$, i.e., $\delta_c\approx 6/4=1.5$. Since $\Delta n_c/\Delta n_a \propto F_a^2/F_c^2$, and $F_{c}$ is several times smaller than $F_a=F_b$, we find that $\Delta n_c/\Delta n_a \sim 10$, and in the first approximation, we may neglect $\Delta n_a$ in Eq.~(\ref{2}). Inserting the above values of $\delta_c$, $n_c$, $F_{c}$ in Eqs.~ (\ref{2}) and (\ref{3}) and using formula (\ref{8b}), we arrive at $(u_{zz}-u_{xx})/B^2\approx 4.8\times 10^{-8}$\ T$^{-2}$ if $|\lambda_{zz}^{(c)} -\lambda_{zz}^{(a)}|=10$ eV (constants $|\lambda_{lk}^{(i)}|$ are of the order of characteristic scale of the electron-band structure in crystals, i.e., $\sim 1$ Ry). The obtained estimate of $(u_{zz}-u_{xx})/B^2$ is close to the value $c_2^z-c_2^x=5.19\times 10^{-8}$ that follows from the data of Fig.~\ref{fig2}. Thus, the anisotropy of the magnetostriction can be understood within the rigid-band approximation.

\subsubsection{Deviation of MS from the $B^2$ law. Existence of small pocket.}

The magnetostriction, like any thermodynamic quantity for a nonmagnetic material with time reversal symmetry, can be represented as a series in powers of $B^2$ for weak magnetic fields when the spacing between Landau levels is less than $T+T_D$ \cite{LL5}. Here $T$ and $T_D$ are the temperature and the Dingle temperature, respectively. For such fields, the oscillating part of the magnetostriction is suppressed, and it is the smooth part of the magnetostriction that is expanded in powers of $B^2$. In the case of the parabolic electron dispersion, this smooth part was calculated  at arbitrary magnetic fields \cite{cichorek}. With this result, we can found the term proportional to $B^4$ in the above-mentioned series, Eq.~(\ref{11b}).
Thus, using formulas (\ref{2}) and (\ref{11b}), we are now  able to calculate the coefficient $c^{z}_{4}-c^{x}_{4}$  characterizing the deviation of $u_{zz}(B)-u_{xx}(B)$ from the quadratic law and to compare it with the value of $c_4^z-c_4^x=-6.97\times 10^{-11}$ T$^{-4}$ derived from the approximation of the experimental data;   Fig.~\ref{fig2}. Neglecting $\Delta n_a$ in Eq.~(\ref{2}) again and using the same $n_c$, $F_c$, $\delta_c$ and $\Lambda_c^{\perp}-\Lambda_c^{\parallel}$, we obtain the value,
\begin{eqnarray*}
c^{z}_{4}-c^{x}_{4}\approx 7.6\times 10^{-14}\ {\rm T}^{-4},
\end{eqnarray*}
which is three order of magnitude smaller than the result extracted from the experimental data. In obtaining this estimate, we have not taken into account that the Fermi energy changes in the magnetic field. This change is proportional to $B^2$, and it introduces a correction of the order of $B^2$ to $c^{z}_{2}-c^{x}_{2}$. This correction means a renormalization of $c^{z}_{4}-c^{x}_{4}$. However, it turns out that the renormalization does not alter the order of magnitude of this quantity, and it cannot explain the above-mentioned huge discrepancy between the theoretical and experimental values of $c^{z}_{4}-c^{x}_{4}$.
The discrepancy can be also due to a deviation of the electron dispersion from the parabolic law assumed in deriving formulas (\ref{3}) and (\ref{11b}). However, again our estimates show that this deviation does not change the order of the magnitude of the $B^2$ and $B^4$ contributions to the magnetostriction. Thus, although existence of the $B^4$ term is quite natural, its anomalously large magnitude is puzzling.

The above considerations argue that a significant deviation of the magnetostriction from the $B^2$ law can be explained only if we assume the existence of a very small charge-carrier pocket of the FS.  When an applied field  exceeds the quantum-oscillation frequency corresponding to this pocket, its  charge carriers are in the ultraquantum limit giving rise to the significant deviation from the quadratic behavior. As inferred from Fig.~\ref{fig3}, such a small pocket can be only due  to the $h3$ holes, although the quantum oscillations associated with this pocket were not observed experimentally \cite{juraszek,xie}. We emphasize, however, that the field-induced contribution of the $h3$ holes to the \textit{anisotropic} part of MS also implies that the rigid-band approximation should fail at least for this group.

\subsubsection{Beyond the rigid-band approximation.}

It was pointed out in Ref.~\cite{khalid} that the three bands $h1$, $h2$, and $h3$ are degenerate at the point $\Gamma$ when the spin-orbit interaction is neglected. With this interaction, the gap $\Delta$ appears that separates the $h1$ and $h2$ bands from the $h3$ holes (Fig.~\ref{fig3}). The dispersion of these three bands is completely analogous to the dispersion  of the valent bands in silicon and germanium, which is described by the well-known three-band model \cite{dresselhaus,kittel}. Within  this model, the effect of the elastic deformations on the hole-bands dispersion was investigated by Pikus and Bir many years ago \cite{pikus,bir}. It turns out that an elastic deformation shifts all the three hole bands in energy by the same value $\lambda_h(u_{xx}+u_{yy}+u_{zz})$ where $\lambda_h$ is a constant of the deformation potential for the holes.
The anisotropy of the deformation, $u_{zz}-u_{xx}$, leads to the splitting of the bands $h1$ and $h2$ at the point $\Gamma$, and this splitting is equal to  $2|(l-m)(u_{zz}-u_{xx})|/3$ where $l$ and $m$ are additional deformation-potential constants characterizing the holes \cite{pikus,bir}. However, the difference  $u_{zz}-u_{xx}$ for LuAs does not exceed $10^{-5}$ at $B< 16$ T (Fig.~\ref{fig2}), and hence the  splitting is less than $0.1$ meV and is negligible as compared to the Fermi energy $\varepsilon_F$ and to the gap $\Delta$ ($|\varepsilon_F|\approx \Delta\approx 0.25$ eV, see Appendix \ref{B}) if $|l-m|\lesssim 1$ Ry in LuAs. We will neglect this very small splitting below, i.e., the gap $\Delta$ is considered to be independent of the deformations. Hence, one might expect the applicability of the rigid-band approximation to the holes in LuAs.
However, the anisotropy of the deformation, $u_{zz}-u_{xx}$, has an effect on the hole masses \cite{pikus,bir}. Moreover, the parameter $\kappa$ defined in Eq.~(\ref{c5}) and specifying this effect is sufficiently large, $\kappa\sim (l-m)/\Delta$. Such sensitivity of the masses to the deformations is due to that in this model, the hole  masses are not determined by $\Delta$. Using these results on the elastic-deformation dependence of the masses, we derive the appropriate formulas for the magnetostriction in Appendix \ref{C}.
Using formulas (\ref{2}), (\ref{8b}), (\ref{c7}), (\ref{c8}), we eventually obtain the following expression for the anisotropic part of the magnetostriction:
\begin{eqnarray}\label{5}
u_{zz}&-&u_{xx}=\frac{\lambda_{zz}^{(c)}- \lambda_{zz}^{(a)}}{C_{33}-C_{13}}(\Delta n_a-\Delta n_c) \nonumber \\
&+&\frac{3}{2(C_{33}-C_{13})}\left(\frac{e}{c}\right)^{5/2}\!
\frac{\kappa(\sqrt{2F_{h1}}+\sqrt{2F_{h2}})\,B^2}{12\pi^2\hbar^{1/2}
|m_{h3}|} \nonumber \\ &-&\frac{3}{2(C_{33}-C_{13})}\frac{\partial\delta\Omega_{h3}}{\partial u_{zz}},
\end{eqnarray}
where the last term is described by Eq.~(\ref{c9}), $\Delta n_i$ are given by Eqs.~(\ref{3}), (\ref{4}), and $\Delta n_a$ can be omitted in the first approximation. This expression is valid at an arbitrary strength of the magnetic field. The first term in Eq.~(\ref{5}) is the contribution of the electron pockets to $u_{zz}-u_{xx}$, whereas the second and third terms describe contributions of the $h1$, $h2$ and $h3$ holes, respectively.

\subsubsection{Estimates of different contributions to $u_{zz}-u_{xx}$.}

To evaluate the relative significance of the terms in Eq.~(\ref{5}), compare the isotropic and anisotropic contributions, e.g., of the pocket $h1$ to the magnetostriction. The isotropic contribution is obtained in the rigid-band approximation and is described by Eqs.~(\ref{19b}) - (\ref{21b}). According to these formulas and Eq.~(\ref{4}), this contribution decreases as $1/\sqrt{F_{h1}}$ with increasing $F_{h1}$, i.e., with increasing the Fermi energy $|\varepsilon_F|$ measured from the band edge for this pocket. On the other hand, the anisotropic contribution of the holes is given by the second term in Eq.~(\ref{5}), and this contribution increases as $\sqrt{F_{h1}}$ with increasing $F_{h1}\propto |\varepsilon_F|$. Then, the ratio of the anisotropic and isotropic contributions is estimated as follows:
 \[
 \frac{e\hbar F_{h1}}{|m_{h3}|c} \frac{\kappa}{\lambda_h}\propto \frac{|\varepsilon_F|}{\Delta}\frac{(l-m)}{\lambda_h},
 \]
where we have used formulas (\ref{16b}) and (\ref{c5}). It is seen that if all the constants of the deformation potential are of the same order, $|\lambda_h|\sim |l-m|$, the contributions become comparable when  $|\varepsilon_F| \sim \Delta$, i.e., when $|\varepsilon_F|$ is of order of the gap in the spectrum. This is  just the case for $h1$ and $h2$ holes (Fig.~\ref{fig3}). In other words, the second term in  Eq.~(\ref{5}), which is associated with the $h1$ and $h2$ holes, can be comparable with their isotropic contribution to the magnetostriction and with the first term in  Eq.~(\ref{5}) since the electron frequencies $F_c$, $F_a$ are not too different from $F_{h1}$, $F_{h2}$.

To estimate the third term in formula (\ref{5}), consider the $h3$ holes (which, according to our assumption, have a small frequency $F_{h3}$) in the ultraquantum regime ($B>F_{h3}$). In this case,  the correct order of the magnitude for the derivative described by Eq.~(\ref{c9}) can be obtained if we replace $F_{h3}$ in formula (\ref{c7}) by $B$. Therefore, in this situation, a value of the third term in Eq.~(\ref{5}) can be estimated as
\begin{eqnarray*}
& &\frac{3}{2(C_{33}-C_{13})}\left(\frac{e}{c}\right)^{5/2}\!
\frac{\sqrt{2}\kappa\,B^{5/2}}{6\pi^2\hbar^{1/2}
|m_{h3}|} \\
&=&\frac{3|n_{h3}|}{8(C_{33}-C_{13})}\left(\frac{e\hbar B}{c|m_{h3}|}\right)\!
\frac{\kappa\,B^{3/2}}{F_{h3}^{3/2}} .
\end{eqnarray*}
Assuming once again that values of the deformation-potential constants are comparable, $|\kappa|\Delta\sim |l-m|\sim |\lambda_{zz}^{(c)}- \lambda_{zz}^{(a)}|$, and taking into account that $|n_{h3}|=n_cF_{h3}^{3/2}/\sqrt{F_a^2F_c}$,
$\Delta \approx |\varepsilon_F|= e\hbar F_{h1}/(|m_{h1}|c)$, we find that the ratio of the third and first terms is approximately equal to $B^{1/2}|m_{h1}|F_c^{3/2}/(|m_{h3}|F_{h1}F_a)\approx 0.022$ at $B=10$ T. Thus, the contribution of the small $h3$ pocket to the $u_{zz}-u_{xx}$ has the magnitude which can explain the observed $10\%$ deviation of MS from the $B^2$ law. However, the magnetic-field dependence of this contribution does not coincide with the simple function $B^4$ used in approximating the experimental data (cf.\ Fig.~\ref{fig2}).

Interestingly, the $h3$ band can manifest itself in the magnetostriction even if $\varepsilon_F+\Delta >0$, i.e., when the $h3$ holes are absent at zero magnetic field and low temperatures. Indeed, if $|\delta_{h3}|>1/2$, and the magnetic field becomes sufficiently large, $B>B_{0+}$, where
  \begin{equation}\label{6}
 \frac{e\hbar B_{0+}}{|m_{h3}|c}= \frac{\varepsilon_F+\Delta}{|\delta_{h3}|-1/2},
  \end{equation}
the edge of the Landau subband $n=0$ raises above the Fermi level, and the $h3$ holes appear in the crystal, see Appendix \ref{B} and  Fig.~\ref{fig4}. For example, if $\delta_{h3}=2$, $|m_{h3}|=0.35m$, and the Fermi level is $1$ meV higher that the upper edge of the $h3$ hole band (i.e., if  $\varepsilon_F+\Delta=1$ meV), we obtain that the $h3$ holes appear at $B>2$ T.

\begin{figure}[tbp] 
 \centering  \vspace{+9 pt}
\includegraphics[scale=0.45]{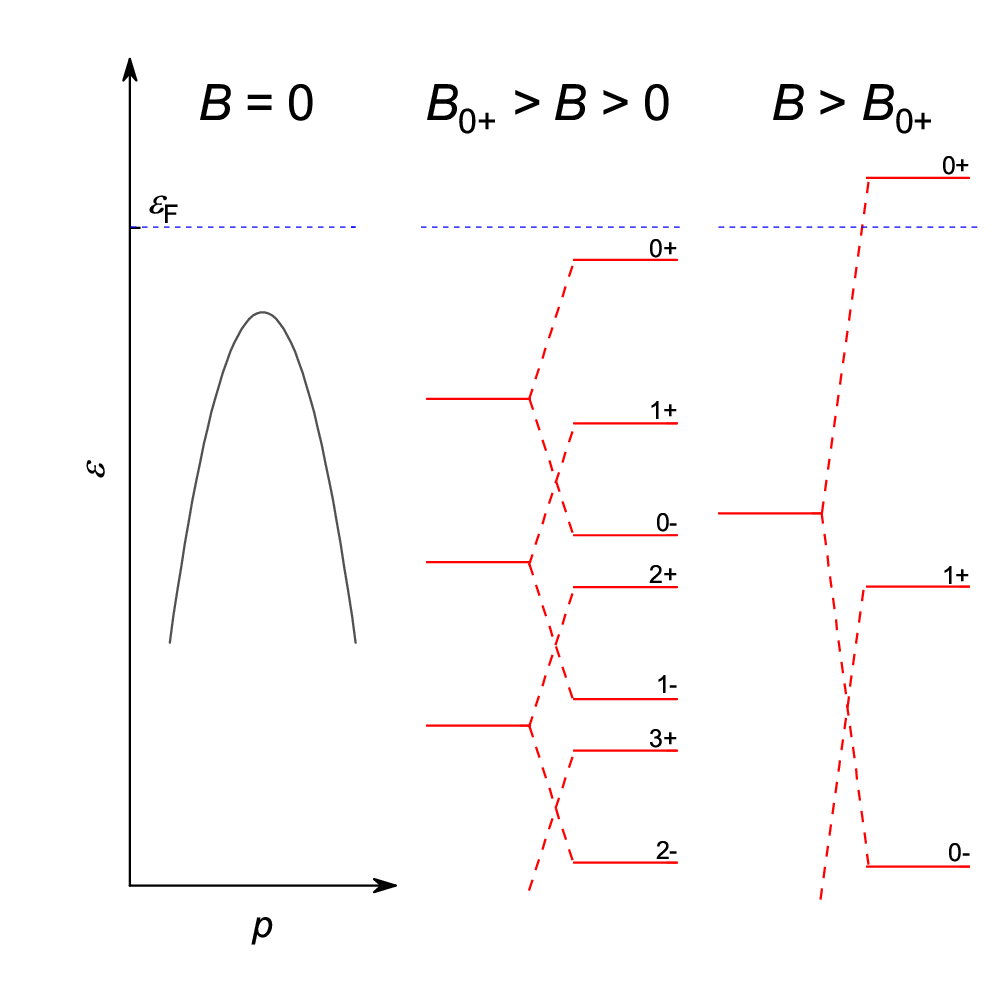}
\caption{\label{fig4} The schematic $B$ dependence of the Landau levels (subbands) of the $h3$ holes when  $1>\delta_{h3}>1/2$. The pairs ($0+$, $0-$),  ($1+$, $1-$), and  ($2+$, $2-$) of the levels result from the splitting of the orbital Landau levels. With increasing $B$, all the spacing between the levels increases, and at $B=B_{0+}$ defined by Eq.~(\ref{6}), the level $0+$ crosses the Fermi energy $\varepsilon_F$. At zero field, $\varepsilon_F$ lies above the edge of the parabolic $h3$ band, as depicted on the left.}
\end{figure}   

\subsection{Isotropic part of the magnetostriction}\label{IIIC}

Before discussing the isotropic part of the magnetostriction, $(u_{zz}+2u_{xx})/3$, consider the redistribution of the charge carriers between the Fermi-surface pockets when the magnetic field is applied to LuAs. This redistribution turns out to be essential for the quantitative description of the isotropic part.

The redistribution follows from the conservation of the charge carriers and is described by the equation that determines $\varepsilon_{F}(B)$, the $B$ dependence of the Fermi energy $\varepsilon_{F}$,
\begin{eqnarray}\label{7}
2\Delta \tilde n_a +\Delta \tilde n_c+\Delta \tilde n_{h1}+\Delta \tilde n_{h2}+\Delta \tilde n_{h3}=0,
\end{eqnarray}
where
 \[
\Delta \tilde n_i\equiv n_i(\varepsilon_F,B)-n_i(\varepsilon_{F0},0) =\nu_i(\varepsilon_{F} -\varepsilon_{F0})+ \Delta n_i,
 \]
$\Delta n_i\equiv n_i(\varepsilon_{F},B)-n_i(\varepsilon_{F},0)$, $\nu_i=\partial n_i(\varepsilon_{F0},0)/\partial\varepsilon_{F0}$ is the density of states for the $i$th pocket, and  $\varepsilon_{F0}\equiv \varepsilon_{F}(B=0)$ is the Fermi energy in absence of the magnetic field. Taking into account that $F_{h3}$ is much less than the other frequencies $F_i$ (and hence $\Delta n_{h3}\gg \Delta n_{i}$ for $i\neq h3$), and that the smallest frequency for the other charge-carrier pockets is $F_c\approx 280$ T,  we may simplify Eq.~(\ref{7}) as follows:
\begin{eqnarray}\label{8}
\nu_{\rm tot}(\varepsilon_F-\varepsilon_{F0})+ \Delta n_{h3} +\Delta n_c\approx 0,
\end{eqnarray}
where $\nu_{\rm tot}=\sum_i \nu_i$ is total density of states, $\Delta n_c$ is given by Eqs.~(\ref{3}), (\ref{4}),  and
$\Delta n_{h3}$ is described by Eq.~(\ref{23b}).
The $B$ dependence of $\varepsilon_{F}$ defined by Eq.~(\ref{8}) means that the frequency $F_{h3}$ gradually changes with $B$, and
using results of Appendix \ref{B}, formula (\ref{8}) can be rewritten in the form:
\begin{eqnarray}\label{9}
 F_{h3}^0&-&F_{h3}\approx -A_h
    {\rm Im}\Biggl \{
    B^{3/2}\sum_{\pm}\zeta(-\frac 12,\,-u_{\pm}+\frac 12 ) \nonumber \\
   &+& i\frac 43 (F_{h3}(1+i\tilde{\Gamma}))^{3/2}
   \Biggr \} -A_e\frac{B^2}{F_c}(\delta_c^2-\frac{1}{12}),
 \end{eqnarray}
that is the equation in the function $F_{h3}(B)$ at given $F_{h3}^0$, $\delta_{h3}$, and $\tilde\Gamma$. Here $F_{h3}^0$ is $F_{h3}$ at $B=0$, $\tilde \Gamma=\pi T_{D,h3}/|\varepsilon_{F0}+\Delta|$, $T_{D,h3}$ is the Dingle temperature for the holes in the band $h3$, $u_{\pm}= (F_{h3}/B)(1+i\tilde\Gamma)\mp \delta_{h3}$, $\zeta(-\frac 12,\,u)$ is the Hurwitz zeta function, and the constants $A_h\approx 3.78\times 10^{-3}\ {\rm T}^{-1/2}$ and $A_e\approx 0.102$ are determined by Eqs.~(\ref{25b}) and (\ref{26b}), respectively.
Since $A_h\sqrt{B}$ and $A_eB/F_c$ are small at $B\le 16$ T, the dependence $F_{h3}(B)$ is sufficiently weak. The other large frequencies $F_i$ can be considered as the parameters that are independent of $B$.
Note that if in Eq.~(\ref{8}), we take into account not only $h3$ holes and the electrons of the $c$ pocket but also the $h1$, $h2$ holes and the electrons in the $a$, $b$ pockets, the coefficient $A_e$ would be multiplied by an additional numerical factor of the order of unity.

With formulas (\ref{6b}), (\ref{7b}) and (\ref{19b}), (\ref{20b}), the combination $u_{zz}+2u_{xx}$ in the symmetric part of the magnetostriction can be written as follows:
\begin{eqnarray}\label{10}
u_{zz}&+&2u_{xx}=(2\Lambda_c^{\perp}+\Lambda_c^{\parallel})(2\Delta \tilde n_a+\Delta \tilde n_c) \nonumber \\
&+&3\Lambda_h(\Delta \tilde n_{h1}+\Delta \tilde n_{h2}+\Delta \tilde n_{h3}) \nonumber \\ &=&(3\Lambda_h-2\Lambda_c^{\perp}-\Lambda_c^{\parallel}) (\Delta \tilde n_{h1}+\Delta \tilde n_{h2}+\Delta \tilde n_{h3}) \nonumber \\
&=&\frac{\lambda_{zz}^{(c)}+2\lambda_{zz}^{(a)}-3\lambda_h}{C_{33} +2C_{13}}(\Delta \tilde n_{h1}+\Delta \tilde n_{h2}+\Delta \tilde n_{h3}) \nonumber \\
&\approx&\frac{\lambda_{zz}^{(c)}+2\lambda_{zz}^{(a)}-3\lambda_h}{C_{33} +2C_{13}}(\nu_h(\varepsilon_F-\varepsilon_{F0})+\Delta n_{h3}) \nonumber \\
&\approx&\frac{\lambda_{zz}^{(c)}+2\lambda_{zz}^{(a)}-3\lambda_h}{C_{33} +2C_{13}}\left (\frac{\Delta n_{h3}}{1+r}-\frac{r \Delta n_c}{1+r} \right ),
\end{eqnarray}
where we have used conservation law (\ref{7}), (\ref{8}), the relation  $\nu_h/\nu_{\rm tot}=r/(1+r)$ between the density of states for the holes $\nu_h$ and $\nu_{\rm tot}$ with $r\approx 0.85$ (see Appendix \ref{B}), and have neglected $\Delta n_{h1}$ and $\Delta n_{h2}$ as compared to $\Delta n_{h3}$. The $\Delta n_{h3}$ and $\Delta n_c$ are described by Eqs.~(\ref{23b}) and (\ref{3}), respectively.  As in the case of Eq.~(\ref{9}), if we take into account the $h1$, $h2$ holes and the electrons of the $a$ and $b$ pockets, this would only change the coefficient before $\Delta n_c$ in the last line of Eq.~(\ref{10}).
Finally, it is important to emphasize that even if the dependence  $\varepsilon_F(B)$ (i.e., $F_{h3}(B)$) is very weak, the term $\nu_h(\varepsilon_F-\varepsilon_{F0})$ associated with the charge-carrier redistribution is always essential in formula (\ref{10}), and it renormalizes $\Delta n_{h3}$. However, terms of such a type are canceled in Eq.~(\ref{5}) for the anisotropic part of the magnetostriction since this formula contains only the difference $\Delta n_a-\Delta n_c$.

It is necessary to note that formulas (\ref{9}), (\ref{10}) are obtained in the rigid-band approximation. We explain in Appendix \ref{C} why the dependence of the masses on deformations (which has been taken into account in Sec.~\ref{IIB}) is unimportant for contributions of the holes to the isotropic part $(u_{zz}+2u_{xx})/3$ of the MS.

\subsection{Comparison of the theoretical and experimental results}

Formula (\ref{10}) shows that the coefficient $\lambda_{zz}^{(c)}+2\lambda_{zz}^{(a)}- 3\lambda_h$ determines the magnitude of $(u_{zz}(B)+2u_{xx}(B))/3$, whereas the shape of this symmetric part of the magnetostriction is described by the sum of  $\Delta n_{h3}(B)$ and the term $\Delta n_c \propto B^2$, with the dependence $\varepsilon_F(B)$ being taken into account. Thus, the fit of the calculated $u_{zz}(B)+2u_{xx}(B)$ to the experimental data can give $F_{h3}^0$, $\delta_{h3}$, and $\lambda_{zz}^{(c)}+ 2\lambda_{zz}^{(a)}-3\lambda_h$. On the other hand, the anisotropic part of the magnetostriction, $u_{zz}-u_{xx}$, permits one to find $\lambda_{zz}^{(c)}- \lambda_{zz}^{(a)}$ and $\kappa$; see Eq.~(\ref{5}).

\begin{figure}[tbp] 
 \centering  \vspace{+9 pt}
\includegraphics[scale=1]{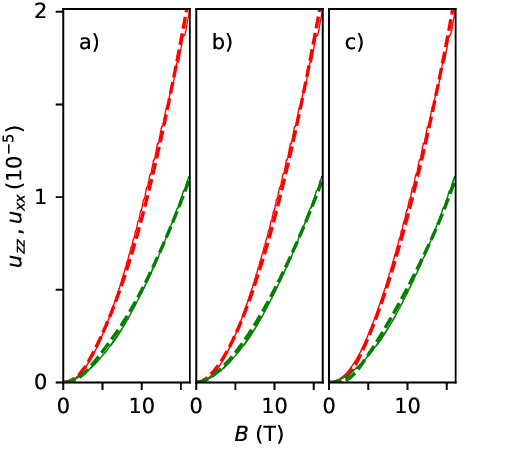}
\caption{\label{fig5} The magnetostriction $u_{zz}(B)$ and $u_{xx}(B)$ calculated with Eqs.~(\ref{3})-(\ref{5}), (\ref{9}), (\ref{10}) for the $a$, $b$, $c$ sets of the parameters in Table \ref{tab1}. The upper and lower curves in all the panels  correspond to $u_{zz}$ and $u_{xx}$, respectively. The solid lines show the experimental data, whereas the dashed lines give the results of the calculations.}
\end{figure}

We calculate $u_{zz}(B)+2u_{xx}(B)$, $u_{zz}(B)-u_{xx}(B)$ with Eqs.~(\ref{3})-(\ref{5}), (\ref{9}), (\ref{10}) and fit the results to the experimental data of Ref.~\cite{juraszek}. The obtained $u_{zz}(B)$ and $u_{xx}(B)$ are shown in Fig.~\ref{fig5}, and the appropriate values of the parameters are presented in Table \ref{tab1}. Note that the fit does not permit us to unambiguously find these values. There are many possible sets of the parameters, and only three of them are presented in Table \ref{tab1}.

It should be also mentioned that we have not tried  to reach a very accurate fit of the calculated curves to the experimental data since we use the simplified model for the dispersion of the $h1$ and $h2$ holes (Appendix \ref{B}), and the electron pockets are considered only in the rigid-band approximation.
Moreover, if we improve the accuracy of the fit, too-large values of the parameters are usually obtained.
However, it is important to emphasize that if we noticeably increase $F_{h3}$ (e.g., above $16$ T), a satisfactory description of the experimental data can never be obtained at reasonable values of the constants of the deformation potential. In other words, in order to reproduce the features of the experimental data pointed out in Sec.~\ref{II}, we should assume existence of a sufficiently small Fermi-surface pocket of the $h3$ holes although this pocket was never detected before.

\begin{table}
\caption{\label{tab1} Three sets of the parameters used in the construction of Fig. \ref{fig5}. The other parameters are: $\Delta=0.25$ eV, $F_c=280$ T, $F_a=F_b=900$ T, $F_{h1}=550$ T, $F_{h2}=1100$ T, $|m_{h1}|/m=0.26$, $|m_{h2}|/m=0.52$, $|m_{h3}|/m=0.35$, $\delta_c=1.5$. The elastic moduli are presented in Table \ref{tab2}. }
\begin{tabular}{c|c|c|c|c|c|c|c}
\hline
\hline \\[-2.5mm]
set&$\varepsilon_F+\Delta$&$F_{h3}$&$\delta_{h3}$  &$T_{D,h3}$&$\lambda_{zz}^{(c)}+2\lambda_{zz}^{(a)}$  &$\lambda_{zz}^{(c)}- \lambda_{zz}^{(a)}$&$\kappa\Delta$ \\
 &  & & & &$-3\lambda_h$ & &  \\
&meV  &T & &K &eV &eV &eV  \\
\colrule
$a$&-0.93 &2.8 &1.0825&1.37&-377.3&-68.5&-30.3 \\ \colrule
$b$&-0.93 &2.8 &2.462 &1.37&-81.8&-64.9&-2.95 \\ \colrule
$c$&1.38    &-   & 2.75    &0.51   &-191&-65.4 & -5.5 \\
\hline \hline
\end{tabular}
\end{table}

\section{Future possible experiments}

Although the above analysis has permitted us to detect a small charge-carrier group in LuAs, we have not been able to unambiguously find all the constants of the deformation potential from the comparison of the experimental and theoretical results. However,
these constants can be found  in special experiments in which the electron ($F_c$, $F_a$) and hole ($F_{h1}$, $F_{h2}$) frequencies are measured in a deformed crystal of LuAs, using the Shubnikov--de Haas or de Haas--van Alphen effects. Namely, it would be sufficient to measure the derivatives $dF_c/dp$, $dF_a/dp$, $dF_{h1}/dp$, and $dF_{h2}/dp$  of these  frequencies with respect to the uniform or uniaxial compression $p$. The appropriate formulas for the analysis of the derivatives are presented in Appendix \ref{D}. If the constants of the deformation potential are determined, the analysis of the magnetostriction will permit one to quantitatively describe the tiny Fermi-surface pocket of the $h3$ holes in LuAs.

The magnetostriction can be used most effectively in studying the electron-band structure of a crystal  when the magnetic field is directed along its highly symmetric axis. In such a case, a part of the charge-carrier pockets of the Fermi surface give equal contributions to the MS, and the number of the parameters of the theory decreases.  In particular, above we have considered the magnetic field parallel to the $4$-fold axis $z$.  If the magnetic field deviates from this axis, the contributions of the electron pockets $a$ and $b$ to the MS are no longer equal, and the above  theory can hardly be used for the initial analysis of the charge-carriers spectra. However, it is worth noting that if the magnetic field is along the direction $[111]$, which is the $3$-fold axis of LuAs, all the electron ellipsoids become equivalent. In this case,  we arrive at $u_{xx}=u_{yy}= u_{zz}$, and the magnetostriction $u_{zz}(B)$ can be described by slightly modified formulas of Sec.~\ref{IIIC}. Thus, measurements $u_{zz}(B)$ at this direction of the magnetic field and a comparison of the obtained data with the appropriate calculations could provide additional information about the spectrum of charge carriers in LuAs.

\begin{figure}[tbp] 
 \centering  \vspace{+9 pt}
\includegraphics[scale=0.52]{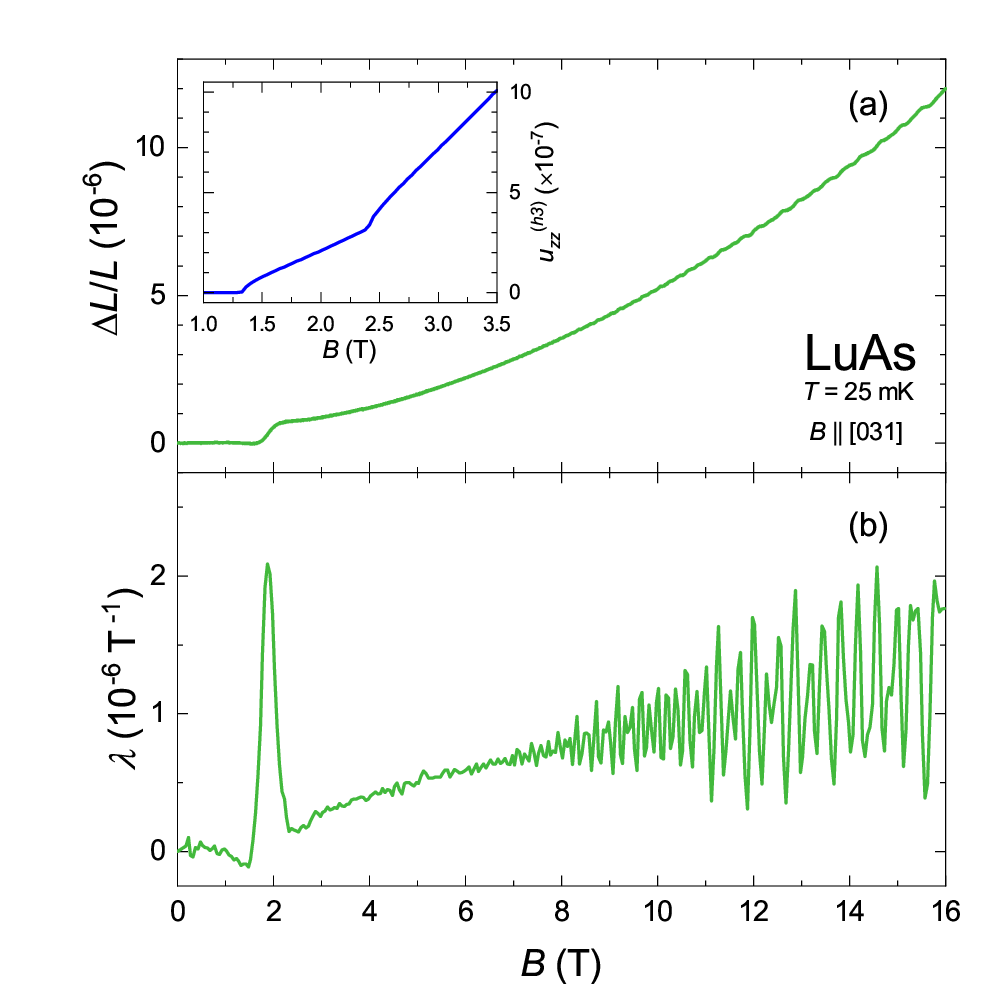}
\caption{\label{fig6} (a) The field-induced relative length change $\Delta L/L$ of LuAs measured along the $z$ axis for $B\parallel [031]$. Inset: Magnetostriction of the $h3$ holes, $u_{zz}^{(h3)}(B)$, calculated with Eqs.~(\ref{19b}), (\ref{20b}), (\ref{23b}) at $\lambda_h=30$ eV, $T_{D,h3}=0.01$ K. Other parameters coincide with those of set $c$ in Table \ref{tab1}. (b) The corresponding magnetostriction coefficient $\lambda(B)$.
} \end{figure}

\begin{figure}[tbp] 
 \centering  \vspace{+9 pt}
\includegraphics[scale=0.53]{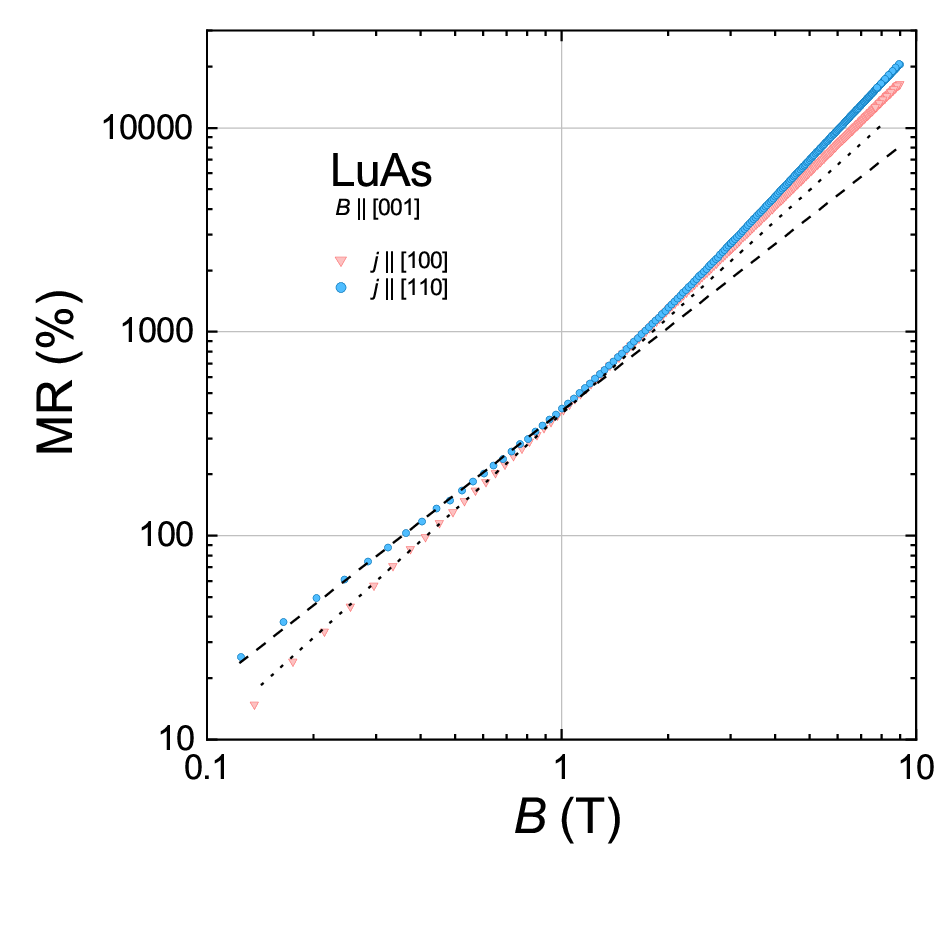}
\caption{\label{fig7} Magnetoresistance MR\,=\,[$R$($B$)\,-\,$R$(0)]/$R$(0) of LuAs measured along the directions $[100]$ (triangles) and $[110]$ (circles) for $B\parallel [001]$ and at $T$\,=\,$2$\,K. The dashed and dotted lines are extrapolation of the experimental data obtained at $B<1$ T to the region of $B>1$ T.
 } \end{figure}

The magnetostriction can be used for investigation of the Lifshitz transitions in metals \cite{lif}. At weak magnetic fields when the spacing between the Landau levels is less that $T+T_D$, changes of the Fermi-surface topology lead to a strong magnetostriction anomaly \cite{lif} that is similar to the anomaly in the thermoelectric power \cite{abr}. The smallness of the $h3$ pocket in LuAs means that this hole group is close to the Lifshitz transition, and the above formulas can actually describe the corresponding  anomaly in MS for strong magnetic fields when the Landau-level spacing exceeds $T+T_D$. The transition occurs when the Landau level $0_+$ crosses the Fermi energy (see Fig.~\ref{fig4}), and the magnetic-field dependence of the MS is expected to exhibit a break at $B=B_{0+}$ (Fig.~\ref{fig6}, Inset). Moreover, it was shown  earlier \cite{m-sh15} that this transition at $T+T_D\to 0$ can manifest itself as a small jump in the magnetostriction, i.e., as the first-order phase transition. In the present work, we have been able to  measure $\Delta L/L$ along the $z$ axis for the magnetic field aligned with the $[031]$ direction. A sharp change in the magnetostriction is observed at about $2$\,T, as depicted in Fig.~\ref{fig6}. Interestingly, at $B$\,$\simeq$\,1\,T, we also observe a break in the slope of the magnetoresistance measured in the $(001)$ plane for $B\parallel [001]$, and a magnitude of this break depends on the direction of the current, as shown in Fig.~\ref{fig7}. Assuming that the observed jump in $u_{zz}$ is due to the transition associated with the $h3$ holes, we can roughly estimated this jump with formulas of Ref.~\cite{m-sh15},
 \[
 \frac{|\Delta \lambda|^3 n_{h3}^2}{(C_{33}+2C_{13})^2(\varepsilon_F+\Delta)}\sim \frac{\Delta \lambda}{\varepsilon_F+\Delta}[u_{zz}(B\sim F_{h3})]^2,
 \]
where $|\Delta \lambda|\sim |\lambda_{zz}^{(c)}+2\lambda_{zz}^{(a)}- 3\lambda_h|/3$. Taking into account that according to Table \ref{tab1}, $|\Delta\lambda|$ is sufficiently large, whereas $\varepsilon_F+\Delta$  is small, we obtain the estimate of the jump $\sim 10^{-7}$ at $u_{zz}(B\sim F_{h3})\sim 10^{-6}$. This very rough estimate is not too different from the experimental value.
However, it is necessary to emphasize that a similar anomaly was not detected when $B$ was along the principal axes of the crystal (cf. Fig.~\ref{fig1}). The $h3$ pocket is spherically symmetric,
and to understand the appearance of the jump only for $B\parallel [031]$, we may assume a sharp dependence of the Dingle temperature $T_{D,h3}$ on the direction of the magnetic field. Namely, for some reason, $T_{D,h3}$ for the $h3$ holes in the ultraquantum regime has to be small when $B$ is close to the direction $[031]$. Another possible explanation is due to that the jump was predicted in Ref.~\cite{m-sh15} within the rigid band approximation, and the spherical symmetry of $u_{zz}^{(h3)}(B)$  also takes place only in this approximation. However, there is a contribution to $u_{zz}$ that results from the deformation dependence of the hole masses. It can be shown that this contribution depends on the direction of the magnetic field and changes its sign as $B$ rotates from the $z$ to $x$ axis. Probably, the jump in the MS appears only at intermediate directions of $B$ when this mass-dependent contribution is small, and the rigid-band approximation works well. It is clear that further experimental and theoretical investigations are required to clarify the origin of the jump in the magnetostriction.

\section{Conclusions}

For a non-magnetic semimetal LuAs, the smooth (nonoscillating in $1/B$) part of the magnetostriction exhibits a noticeable deviation from the quadratic-in-field dependence. This universal $B^2$ law must be observed in applied magnetic fields essentially smaller than the fundamental frequencies $F_i$ of the quantum oscillations. We show that for LuAs with its $F_i\ge 280$\,T, the above deviation at $B\le 16$ T is caused by a tiny hole group. Although the band-structure calculations are indicative of a possible existence of this group,  it was not detected in magnetic oscillations before. Thus, we do demonstrate how the dilatometry permits a discovery of the small Fermi-surface pocket. Using the developed theory of the magnetostriction, we also quantitatively describe the magnetic-field dependence of the magnetostriction. Measurements of the magnetostriction and of pressure dependences of the observed frequencies $F_i$ could significantly extend possibilities of studying the tiny Fermi-surface pocket.  We believe that similar studies can result in a better understanding of other systems with small groups of charge carriers, in particular, of topological semimetals.

\begin{acknowledgments}
  Yu.V.S. acknowledges the program for scientists from Ukraine 2022/01/3/ST3/00083 by the National Science Centre, Poland.
\end{acknowledgments}

\appendix

\section{Magnetostriction of crystals with cubic symmetry}\label{A}

Formulas for the magnetostriction can be derived by a minimization of the $\Omega$ potential for the deformed crystal placed in the external magnetic field $B$ with respect to the strain tensor $u_{ik}$.
In the case of a crystal with the cubic structure,  this $\Omega$ potential has the form:
\begin{eqnarray}\label{1a}
\Omega&=&\frac{1}{2}C_{33}(u_{xx}^2+u_{yy}^2+u_{zz}^2)
\nonumber \\
&+&C_{13}(u_{xx}u_{zz}+u_{yy}u_{zz}+u_{xx}u_{yy})\nonumber \\&+& 2C_{66}(u_{xy}^2+u_{xz}^2+u_{yz}^2) \nonumber \\
&+&\Delta \Omega_{\rm el}(u_{ik},B)- \Delta \Omega_{\rm el}(u_{ik},0),
\end{eqnarray}
where $C_{mn}$ are the elastic moduli of the crystal \cite{LL-sl}, $\Delta \Omega_{\rm el}(u_{kl},B)\equiv
\Omega_{\rm el}(u_{kl},B)-\Omega_{\rm el}(0,B)$, and $\Omega_{\rm el}(u_{kl},B)=\sum_{i=1}^N \Omega_{i}(u_{kl},B)$ is the deformation-dependent part of the electron $\Omega$ potential for the crystal with $N$ pockets of charge carriers. Apart from the deformation, this part depends on the magnetic induction  $B$,  the temperature $T$, and the chemical potential $\zeta$ (we do not indicate explicitly $T$ and $\zeta$ here). The first three terms in Eq.~(\ref{1a}) give the total elastic energy of a deformation. This energy is partly produced by  $\Delta \Omega_{\rm el}(u_{kl},0)$, and hence the difference of the elastic terms and $\Delta \Omega_{\rm el}(u_{kl},0)$ is the elastic energy that is not associated with the Fermi-surface pockets under study. The term $\Delta \Omega_{\rm el}(u_{kl},B)$ describes the total change in the $\Omega$ potential of these pockets in the magnetic field under the deformation.
The last two terms in Eq.~(\ref{1a}) can be also rewritten as follows:
 \begin{eqnarray*}
\Delta \Omega_{\rm el}(u_{ik},B)&-& \Delta \Omega_{\rm el}(u_{ik},0)=
[\Omega_{\rm el}(u_{kl},B)-\Omega_{\rm el}(u_{kl},0)] \nonumber \\ -[\Omega_{\rm el}(0,B)&-&\Omega_{\rm el}(0,0)]=
u_{kl}\frac{\partial [\delta \Omega_{\rm el}(B)]}{\partial u_{kl}},
 \end{eqnarray*}
where $\delta \Omega_{\rm el}(B)\equiv \Omega_{\rm el}(u_{kl}\to 0,B)-\Omega_{\rm el}(u_{kl}\to 0,0)$. In particular, for the case of the weak magnetic fields, $\delta \Omega_{\rm el}(B)=-\chi B^2/2$ where $\chi$ is the magnetic susceptibility produced by the Fermi-surface pockets under study.

The minimization of Eq.~(\ref{1a}) gives the set of the equations in the tensor  $u_{kl}$, which defines the magnetostriction, i.e., the deformation of the crystal in the magnetic field,
\begin{eqnarray}\label{2a}
C_{33}u_{zz}&+&C_{13}(u_{xx}+u_{yy})=-\frac{\partial [\delta \Omega_{\rm el}(B)]}{\partial u_{zz}}, \nonumber \\
C_{33}u_{xx}&+&C_{13}(u_{zz}+u_{yy})=-\frac{\partial [\delta \Omega_{\rm el}(B)]}{\partial u_{xx}}, \nonumber \\
C_{33}u_{yy}&+&C_{13}(u_{zz}+u_{xx})=-\frac{\partial [\delta \Omega_{\rm el}(B)]}{\partial u_{yy}}, \nonumber \\
4C_{66}u_{xy}&=&-\frac{\partial [\delta \Omega_{\rm el}(B)]}{\partial u_{xy}}, \\
4C_{66}u_{xz}&=&-\frac{\partial [\delta \Omega_{\rm el}(B)]}{\partial u_{xz}}, \nonumber \\
4C_{66}u_{yz}&=&-\frac{\partial [\delta \Omega_{\rm el}(B)]}{\partial u_{yz}}. \nonumber
\end{eqnarray}
With these equations, one can derive that all off-diagonal $u_{lk}$ (when $l\neq k$) are equal to zero for crystals with the cubic (tetragonal, orthogonal) symmetry. Indeed, consider the situation when $u_{l\neq k}\equiv 0$, and only the diagonal components of the strain tensor differ from zero. In this case, the deformed crystal has the orthogonal symmetry, and $(\partial[\delta \Omega_{\rm el}]/\partial\theta)_{\theta=0}=0$ where $\theta$ is the angle of a deviation of ${\bf B}$ from one of the coordinate axes $x_i$ ($x_i=x,y,z$), which can be named as the principal directions of $\delta \Omega_{\rm el}({\bf B})$. Hence, $\delta \Omega_{\rm el}(B, \theta)-\delta \Omega_{\rm el}(B,\theta=0)\propto \theta^2$ if such a deviation occurs.
At nonzero $u_{l\neq k}$, these principal directions of $\delta \Omega_{\rm el}({\bf B})$ depart from the coordinate axes by the angles of the order of $u_{ik}$. Therefore, for the magnetic field parallel to a coordinate axis, $\delta\Omega_{\rm el}({\bf B})\propto (u_{l\neq k})^2$, and the last three formulas in Eqs.~(\ref{2a}) give $u_{xy}=u_{xz}=u_{yz}=0$.

\section{Magnetostriction in the rigid-band approximation} \label{B}

Within the rigid-band approximation, an elastic deformation shifts the electron bands as a whole and does not change their dispersion. In this approximation, a shift $\Delta\varepsilon_i$ of the $i$th electron energy band $\varepsilon_i({\bf p})$ under the deformation coincides with the shift of its edge and is proportional to the strain tensor $u_{kl}$, $\Delta\varepsilon_i= \sum_{k,l}\lambda_{kl}^{(i)}u_{kl}$, where ${\bf p}$ is the electron quasimomentum, and $\lambda_{kl}^{(i)}$ are the constants of the deformation potential for the $i$th band \cite{abr}. Therefore, in the minimization of $\Omega$ potential, one can use the following relation:
\[
\frac{\partial\Omega_{i}(u_{kl},B)}{\partial u_{kl}}= -\frac{\partial\Omega_{i}(\zeta-\varepsilon_i,B)}{\partial \zeta} \frac{\partial\Delta\varepsilon_i}{\partial u_{kl}}=\lambda_{kl}n_i(B),
\]
where $n_i(B)=-\partial\Omega_{i}(\zeta,B)/\partial \zeta$ is the density of charge carriers of the pocket $i$ in the magnetic field. Then, equations (\ref{2a}) reduce to the following form:
 \begin{eqnarray}\label{1b}
C_{33}u_{zz}\!&+&\!C_{13}(u_{xx}\!+\!u_{yy})\!= \!-\sum_i\lambda_{zz}^{(i)}\Delta n_i, \nonumber \\
C_{33}u_{xx}\!&+&\!C_{13}(u_{zz}\!+ \!u_{yy})\!=\!-\sum_i\lambda_{xx}^{(i)}\Delta n_i, ~~~\\
C_{33}u_{yy}\!&+&\!C_{13}(u_{zz}\!+\!u_{xx})\!= \!-\sum_i\lambda_{yy}^{(i)}\Delta n_i, \nonumber
 \end{eqnarray}
where $\Delta n_i\equiv n_i(B)-n_i(0)$ is the change in the charge-carrier density of the pocket $i$ in the magnetic field for the undeformed crystal, and  $i=1,\dots, N$ usually runs several groups of the equivalent Fermi pocket. (We call pockets equivalent if, in absence of the magnetic field, they transform into each other under symmetry operations.) In Eqs.~(\ref{1b}), we have taken into account that $u_{lk}=0$ if $l\neq k$. Solving this set of the linear  equations, one obtains formulas for $u_{xx}$, $u_{yy}$, and $u_{zz}$. Note that these formulas are further simplified  when a symmetry of the Fermi-surface pockets (i.e., the symmetry of the constants $\lambda_{ll}^{(i)}$)  is taken into account (see below). The elastic moduli of LuAs were calculated in Ref.~\cite{mir} (Table \ref{tab2}). Thus, the magnetostriction can be easily found when $\Delta n_i$ are  known.

\begin{table}
\caption{\label{tab2}\textbf{The elastic moduli of LuAs \cite{mir}}.}
\begin{tabular}{c|c|c}
\hline
\hline \\[-2.5mm]
$C_{33}$ (GPa)  &$C_{13}$ (GPa)  &$C_{66}$ (GPa)  \\
\colrule
$188.6$ & $25.0$ & $43.7$ \\
\hline \hline
\end{tabular}
\end{table}

\subsection{Electrons in LuAs}

The electron pockets are located at the $X$ points of the four-fold rotation axes $a$, $b$, $c$ (Fig.~2). It follows from symmetry considerations that for the electron pocket lying in the $i$th axis ($i=a$, $b$, $c$), the parameters $\lambda_{kl}^{(i)}$ of the deformation potential satisfy the following relations:
\begin{eqnarray}\label{2b}
\lambda_{xx}^{(a)}=\lambda_{zz}^{(c)},\ \ \lambda_{xx}^{(b)}=\lambda_{xx}^{(c)}=\lambda_{zz}^{(a)},
 \nonumber \\
\lambda_{yy}^{(a)}=\lambda_{yy}^{(c)}=\lambda_{zz}^{(a)},
\ \ \lambda_{yy}^{(b)}=\lambda_{zz}^{(c)}, \nonumber \\
\lambda_{zz}^{(b)}=\lambda_{zz}^{(a)}, \\
\lambda_{k\neq l}^{(a)}=\lambda_{k\neq l}^{(b)}=\lambda_{k\neq l}^{(c)}=0. \nonumber
\end{eqnarray}
In other words, we have only two independent constants $\lambda_{zz}^{(c)}$ and  $\lambda_{zz}^{(a)}$ that determine all other $\lambda_{kl}^{(i)}$ for the electrons. The dispersion relations for the electrons near the point X are also specified by the two effective masses $m_c$ and $m_a$. Taking into account the deformation shift of the electron pockets, these relations looks as follows:
\begin{eqnarray}\label{3b}
\varepsilon^{(c)}\!({\bf p})\!=\!\frac{p_z^2}{2m_c}\!+\! \frac{p_x^2\!+\!p_y^2}{2m_a}\!+\! \lambda_{zz}^{(c)}u_{zz}\!+\!\lambda_{zz}^{(a)}(u_{xx}\!+\!u_{yy}),~~\\
\varepsilon^{(a)}\!({\bf p})\!=\!\frac{p_x^2}{2m_c}\!+\! \frac{p_z^2\!+\!p_y^2}{2m_a}\!+\! \lambda_{zz}^{(c)}u_{xx}\!+\!\lambda_{zz}^{(a)}(u_{zz}\!+\!u_{yy}),~~ \label{4b}
\\
\varepsilon^{(b)}\!({\bf p})\!=\!\frac{p_y^2}{2m_c}\!+\! \frac{p_z^2\!+\!p_x^2}{2m_a}\!+\! \lambda_{zz}^{(c)}u_{yy}\!+\!\lambda_{zz}^{(a)}(u_{zz}\!+\!u_{xx}).~~
\label{5b}
\end{eqnarray}
As to the electron contribution to the magnetostriction, we find from Eqs.~(\ref{1b}) and (\ref{2b}) that
\begin{eqnarray}\label{6b}
u_{zz}^{(e)}&=&\Lambda_a^{\parallel}(\Delta n_a+ \Delta n_b)+\Lambda_c^{\parallel}\Delta n_c, \nonumber \\
u_{xx}^{(e)}&=&u_{yy}^{(e)}=\Lambda_a^{\perp}(\Delta n_a+ \Delta n_b)+\Lambda_c^{\perp}\Delta n_c,
\end{eqnarray}
where
\begin{eqnarray}\label{7b}
\Lambda_a^{\parallel}&=&\Lambda_c^{\perp}=\frac{\lambda_{zz}^{(c)}C_{13}- \lambda_{zz}^{(a)}C_{33}}{C_{33}(C_{33}+C_{13})-2(C_{13})^2},
 \nonumber \\
\Lambda_c^{\parallel}&=&-\frac{\lambda_{zz}^{(c)}(C_{33}+C_{13})- 2\lambda_{zz}^{(a)}C_{13}}{C_{33}(C_{33}+C_{13})-2(C_{13})^2},  \\
\Lambda_a^{\perp}&=&-\frac{\lambda_{zz}^{(c)}C_{33}-\lambda_{zz}^{(a)} (2C_{13}-C_{33})}{2[C_{33}(C_{33}+C_{13})-2(C_{13})^2]}= \frac{\Lambda_c^{\parallel}+\Lambda_c^{\perp}}{2}.
\nonumber
\end{eqnarray}
Formula (\ref{2}) with
\begin{eqnarray}\label{8b}
\Lambda_c^{\perp}&-&\Lambda_c^{\parallel}= \frac{\lambda_{zz}^{(c)}- \lambda_{zz}^{(a)}}{C_{33}-C_{13}} \\
 &\approx&0.54\times 10^{-23}[{\rm cm}^3/{\rm eV}]\times (\lambda_{zz}^{(c)}- \lambda_{zz}^{(a)}), \nonumber
\end{eqnarray}
follows from Eqs.~(\ref{7b}) at $\Delta n_a=\Delta n_b$. In Eq.~(\ref{8b}), we have also used  the data of Table \ref{tab2}.

At $u_{zz}=u_{xx}=u_{yy}=0$, the spectrum in a magnetic field $B\parallel z$ for the electrons described by Eqs.~(\ref{3b})-(\ref{5b}) is well known,
\begin{eqnarray}\label{9b}
 \varepsilon^{(c)}(n,p_z)&=&\frac{e\hbar B}{cm_a}(n\pm\delta_c)+\frac{p_z^2}{2m_c}, \\
 \varepsilon^{(a)}(n,p_z)=\varepsilon^{(b)}(n,p_z)&=&\frac{e\hbar B}{c(m_am_c)^{1/2}}(n\pm\delta_a)+\frac{p_z^2}{2m_a}, \nonumber
\end{eqnarray}
where $n=0, 1, \dots$, $\delta_c=g_cm_a/(4m)$, $\delta_a=g_a(m_am_c)^{1/2}/(4m)$, $m$ is the free electron mass, $g_c$ and $g_a$ are the \textit{g} factors of the electrons in the pockets $c$ and $a$, respectively, at $B\parallel z$. With these spectra, one can find $\Delta n_a$, $\Delta n_b$, and $\Delta n_c$.  In particular, for the weak magnetic fields ($B \ll F_c$, $F_a$), we have \cite{cichorek}:
\begin{eqnarray}\label{10b}
 \Delta n_c&=&\frac{3nB^2}{8F_c^2}(\delta_c^2-\frac{1}{12}), \\
 \Delta n_a&=&\Delta n_b=\!\frac{3nB^2}{8F_a^2}(\delta_a^2-\frac{1}{12}), \nonumber
 \end{eqnarray}
where $n=n(0)$ is the density of the electrons per one pocket at $B=0$, and $F_c\approx 280$ T, $F_a\approx 900$ T \cite{xie} are the frequencies of the quantum oscillation produced by the ellipsoids $c$ and $a$, respectively, at $B\parallel z$. With Eqs.~(\ref{6b}), (\ref{7b}), and (\ref{10b}), we obtain the magnetostriction produced by the small electron pockets only in the weak magnetic fields. However, due to the large values of $F_c$ and $F_a$, all the fields accessible in experiments are weak for the electrons in LuAs. Nevertheless, for reference, we give the correction to formulas (\ref{10b}), which  is proportional to $B^4$ and can be derived from results of Ref.~\cite{cichorek},
 \begin{eqnarray}\label{11b}
\delta(\Delta n_i)=\frac{n_iB^4}{F_i^4}\left( \frac{240\delta_i^4- 120\delta_i^2+7}{10240}\right).
 \end{eqnarray}

\subsection{The holes in LuAs}

In LuAs, the centers of all the hole pockets $h1$, $h2$, $h3$  are at the point $\Gamma$ of its  Brillouin zone, and the appropriate three holes bands are close to each other at the point $\Gamma$ (Fig.~2).
The dispersion of these hole bands in the vicinity of the point $\Gamma$, in fact, was obtained from the three-band model \cite{dresselhaus,kittel} and has the following form:
\begin{eqnarray}\label{12b}
\varepsilon_{h1,h2}({\bf p})&=&\frac{F+G}{2}\pm \sqrt{\frac{(F-G)^2}{4}+|H|^2+|I|^2},~~~  \\
\varepsilon_{h3}({\bf p})&=&\frac{F+G}{2}-\Delta, \label{13b}
\end{eqnarray}
where the energy is measured from $\varepsilon_{h1}({\bf p}=0)= \varepsilon_{h2}({\bf p}=0)$,
\begin{eqnarray}\label{14b}
F&=&\frac{L+M}{2}(p_x^2+p_y^2)+Mp_z^2, \nonumber\\
G&=& \frac{M+2L}{3}p_z^2+\frac{L+5M}{6}(p_x^2+p_y^2), \nonumber \\
\frac{F+G}{2}&=&\frac{L+2M}{3}(p_x^2+p_y^2+p_z^2)\equiv   \frac{p^2}{2m_{h3}}, \\
H&=&-\frac{N}{\sqrt{3}}(p_yp_z+ip_x p_z), \nonumber \\
I&=&\frac{1}{\sqrt{12}}[(L-M)(p_x^2-p_y^2)-2iNp_yp_x], \nonumber
\end{eqnarray}
$L$, $M$, $N$ are constants of the order of $1/m$, and $\Delta$ is the gap separating the band $h3$ at the point $\Gamma$ from the bands $h1$ and $h2$ which are degenerate at this point (Fig~2). Within the rigid-band approximation, the deformation shift $\Delta \varepsilon_i$ for all the hole bands is the same and is equal to $\lambda_h(u_{xx}+ u_{yy}+ u_{zz})$ \cite{pikus,bir}. Note that this shift is determined by a single constant $\lambda_h$ of the deformation potential, and the shift does not change the gap $\Delta$ between the bands $h1$, $h2$ and $h3$ at the point $\Gamma$.

Due to the square root in formula (\ref{12b}), the Fermi-surface  pockets $h1$ and $h2$ look like two corrugated spheres. Even for the case of a weak magnetic field, a calculation of the magnetic susceptibility produced by such pockets (and hence of the corresponding  $\Delta n_i$) is a complicated problem. Below we shall consider these pockets as usual spheres, i.e., we shall use the following simple model for the bands $\varepsilon_{h1}({\bf p})$ and $\varepsilon_{h2}({\bf p})$,
 \begin{eqnarray}\label{15b}
\varepsilon_{h1,h2}({\bf p})=\frac{p^2}{2m_{h1,h2}},
\end{eqnarray}
but we choose the Fermi energy $\varepsilon_F$ and the effective masses $m_{h1}$, $m_{h2}$ so that they reproduce the parameters extracted from the experimental data. (Note that all $m_{hi}$ and $\varepsilon_F$ are negative here.) In particular, formulas (\ref{12b}) - (\ref{14b}) enable one to calculate the areas $S_{hi}$ and the cyclotron masses $m_{hi}$ of the extremal cross sections of the three holes pockets by the plane $p_z=0$. As a result, we obtain the following simple  relations (see  formula (37) in Supplementary Materials to Ref.~\cite{cichorek} and Ref.~\cite{m-sh21a}):
\begin{eqnarray}\label{16b}
\frac{S_{h1}}{2\pi m_{h1}}&=&\frac{e\hbar F_{h1}}{m_{h1}c}=\frac{S_{h2}}{2\pi m_{h2}}=\frac{e\hbar F_{h2}}{m_{h2}c}=\varepsilon_F,  \\
\frac{S_{h3}}{2\pi m_{h3}}&=&\frac{e\hbar F_{h3}}{m_{h3}c}= \varepsilon_F+\Delta .
 \label{17b}
\end{eqnarray}
In fact, formulas (\ref{16b}) and (\ref{17b}) are one and the same expression since $\varepsilon_F+\Delta$ is the Fermi energy measured from the top, $-\Delta$, of the hole band $h3$. Knowing the frequency of the quantum oscillations for the holes $h1$, $F_{h1}\approx 550$ T, and the appropriate  mass $|m_{h1}|/m\approx 0.26$ \cite{xie}, we find $|\varepsilon_F| \approx 250$ meV. However, the Fermi level is near the edge of the $h3$ band \cite{xie}, and so $\Delta\approx 250$ meV. With this value of $\varepsilon_F$, we find $|m_{h2}|/m\approx 0.52$ for the oscillations corresponding to the frequency $F_{h2}\approx 1100$ T of  the $h2$ holes \cite{xie}. A comparison Eqs. ~(\ref{12b}) and (\ref{13b}) also indicates that $[\varepsilon_{h1}({\bf p})+\varepsilon_{h2}({\bf p})]/2=\varepsilon_{h3}({\bf p})+\Delta$, and so we take
\begin{eqnarray}\label{18b}
\frac{1}{m_{h3}}=\frac{1}{2}\left (\frac{1}{m_{h1}}+ \frac{1}{m_{h2}}\right),
\end{eqnarray}
i.e., $|m_{h3}|/m\approx 0.35$.

Solving Eqs.~(\ref{1b}), we obtain the following expression for the contribution of the hole pocket $hi$ ($hi=h1,h2,h3$) to the magnetostriction of LuAs:
\begin{eqnarray}\label{19b}
u_{zz}^{(hi)}&=&u_{xx}^{(hi)}=u_{yy}^{(hi)}=\Lambda_h\Delta n_{hi}, \end{eqnarray}
where
\begin{eqnarray}\label{20b}
\Lambda_h=-\frac{\lambda_h}{ C_{33}+2C_{13}}.
\end{eqnarray}
For the weak magnetic fields ($B\ll F_{hi}$), the $\Delta n_{hi}$ is  described by the formula similar to Eqs.~(\ref{10b}),
\begin{eqnarray}\label{21b}
 \Delta n_{hi}&=&\frac{3n_{hi}B^2}{8F_{hi}^2}(\delta_{hi}^2-\frac{1}{12}), \end{eqnarray}
where $n_{hi}=n_{hi}(0)$ is density of the holes in the pocket $hi$ at the zero magnetic field, $\delta_{hi}=g_{hi}m_{hi}/(4m)$, and $g_{hi}$ is the appropriate \textit{g} factor. It is necessary to emphasize  that we define the densities of the holes as negative quantities, $n_{hi}<0$, whereas the electron density is implied to be positive, $n>0$. For the pocket $h3$, which exists if $\varepsilon_F<-\Delta$, the frequency $F_{h3}$ is relatively small, and the holes of this pocket can be in the ultraquantum regime when $B>F_{h3}$. In this case, $\Delta n_{h3}(B)$ can be described by the formulas presented in Supplementary Materials to Ref.~\cite{cichorek},
\begin{eqnarray} \label{22b}
    \Delta n_{h3}&=&\frac{(e/c)^{3/2}}{\sqrt{2}\pi^2\hbar^{3/2}} \Biggl \{ -B^{3/2}\!\!\sum_{n,\pm}
	( \frac{F_{h3}}B -n - \frac 12 \mp \delta_{h3})^{1/2} \nonumber \\
    &+& \frac 43 F_{h3}^{3/2} \Biggr \},
\end{eqnarray}
where the summation is carried out over those integer $n\ge 0$ and those signs before $\delta_{h3}$ for which the summands are positive. This formula is obtained from the expression that is valid at arbitrary $B$ and a nonzero Dingle temperature \cite{cichorek},
\begin{eqnarray} \label{23b}
    \Delta n_{h3}&=&\frac{3|n_{h3}|}{4}
    {\rm Im}\Biggl \{ \frac{B^{3/2}}{F_{h3}^{3/2}}\sum_{\pm}\zeta(-\frac 12,\, -u_{\pm} + \frac 12 ) \nonumber \\
   &+& i\frac 43 (1+i\tilde{\Gamma})^{3/2} \Biggr \},
  \end{eqnarray}
where $|n_{h3}|=(3\sqrt{\pi})^{-1}(8F_{h3}^3/\phi_0^3)^{1/2}$, $\phi_0$ is the flux quantum, $\tilde \Gamma=\pi T_{D,h3}/|\varepsilon_F+\Delta|$, $T_{D,h3}$ is the Dingle temperature for the holes in the band $h3$, $u_{\pm}= (F_{h3}/B)(1+i\tilde\Gamma)\mp \delta_{h3}$, and  $\zeta(-\frac 12,\,u)$ is the Hurwitz zeta function \cite{batem}.
Note that according to Eq.~(\ref{17b}), $F_{h3}/B=m_{h3}c(\varepsilon_F+\Delta)/e\hbar B$, and for $\delta_{h3}>1/2$, the sums in Eqs.~(\ref{22b}) and (\ref{23b}) can contain nonzero terms even when $\varepsilon_F+\Delta>0$ and  $F_{h3}/B<0$, see Eq.~(\ref{6}).

Thus, within the rigid-band approximation, the holes give the isotropic contribution (\ref{19b}) to the magnetostriction, whereas the electron contribution (\ref{6b}) can explain its anisotropy, i.e., the nonzero values of $u_{zz}(B)-u_{xx}(B)$, Fig.~\ref{fig1}.

\subsection{Dependence of $\varepsilon_F$ on $B$}\label{B3}

Equation (\ref{8}) in the main text describes the $B$-dependent shift of the Fermi energy $\varepsilon_{F}$ relative to its value at $B=0$, $\varepsilon_{F0}\equiv \varepsilon_F(B=0)$.
To proceed, it is convenient to express $\nu_{\rm tot}$ in that equation in terms of $|n_{h3}(\varepsilon_{F0})|/|\varepsilon_{F0}+\Delta|$.

We have found in this Appendix  that $|\varepsilon_{F0}|\approx  250$ meV for the $h1$ and $h2$ holes. As to the Fermi level measured from the bottom of the electron band, it is equal to $\varepsilon_F^{e}\approx 210$ meV.  This estimate follows from  $F_c\approx 280$ T, the appropriate  cyclotron mass $m_c\approx 0.155m$  \cite{xie}, and a formula similar to Eq.~(\ref{17b}). Therefore, we have $r\equiv \varepsilon_F^{ e}/|\varepsilon_{F0}|\approx 0.85$. According to Refs.~\cite{cichorek,xie}, the electron ($n_e=3n_c$) and hole ($|n_h|=|n_{h1}|+|n_{h2}| +|n_{h3}|$) densities are equal to each other with an accuracy about $1-2\, \%$. Then, taking into account that $\nu_{h1}+\nu_{h2}\gg \nu_{h3}$, $n_c/|n_{h3}|=(F_cF_a^2/F_{h3}^3)^{1/2}$ [see Eq.~(\ref{4})], and  $\varepsilon_F^{\rm e}/|\varepsilon_{F0}+\Delta|=F_c|m_{h3}|/(F_{h3}m_c)$ [see Eq.~(\ref{17b})], we arrive at
 \begin{eqnarray}\label{24b}
 \nu_{\rm tot}&\approx& \frac{3}{2}\frac{3n_c}{\varepsilon_F^e}+
\frac{3}{2}\frac{|n_{h1}|+|n_{h2}|}{|\varepsilon_{F0}|}\approx  \frac{9n_c}{2\varepsilon_F^e}(1+r) \nonumber \\ &\approx& \frac{3|n_{h3}|}{4|\Delta+\varepsilon_{F0}|}\frac{1}{A_hF_{h3}^{1/2}},
 \end{eqnarray}
where
  \begin{equation}\label{25b}
  A_h\equiv \frac{F_c^{1/2}|m_{h3}|}{6(1+ r)F_am_c}\approx 3.78\times 10^{-3}\ {\rm T}^{-1/2}
  \end{equation}
for $F_c=280$ T, $F_a=900$ T, $|m_{h3}|=0.35m$, $m_c=0.155m$. These formulas show that if, e.g., $F_{h3}=4$ T, the total density of states is approximately $65$ times larger than $\nu_{h3}= (3/2)|n_{h3}|/|\Delta+\varepsilon_{F0}|$.

The $B$ dependence of $\varepsilon_F$ leads to the dependence of the frequency $F_{h3}$ on $B$ since $(F_{h3}^0-F_{h3}(B))/F_{h3}^0= (\varepsilon_F-\varepsilon_{F0})/|\Delta+\varepsilon_{F0}|$ where $F_{h3}^0\equiv F_{h3}(B=0)$. Then,
inserting formulas (\ref{23b}), (\ref{24b}), and expression (\ref{10b}) for $\Delta n_c$ into Eq.~(\ref{8}), we arrive at Eq.~(\ref{9}) with
 \begin{equation}\label{26b}
  A_e=\frac{1}{12(1+r)}\frac{|m_{h3}|}{m_c}\approx 0.102.
   \end{equation}

\section{Magnetostriction beyond the rigid-band approximation} \label{C}

Considering the holes bands as an example, let us now go beyond the rigid-band approximation.
Using the three-band model \cite{dresselhaus,kittel}, an effect of the elastic deformation on the dispersion of the hole bands was studied in Ref.~\cite{pikus,bir} beyond the rigid-band approximation. It turned out that this effect is determined by the three constants of the deformation potential (namely, by $l, m, n$), and in these notations, one has $\lambda_h=(l+2m)/3$ for $\lambda_h$ introduced above. The results of Pikus and Bir \cite{pikus,bir} also show that the uniform compression (when $u_{xx}=u_{yy}=u_{zz}\neq 0$) leads only to the shift of the three hole bands as whole, but it does not change their dispersion. However, under an arbitrary deformation, the hole band $h3$ changes as follows:
\begin{eqnarray} \label{c1}
    \varepsilon_{h3}({\bf p})&\approx&\frac{p^2}{2m_{h3}}-\Delta  +\lambda_h(u_{xx}+u_{yy}+u_{zz}) \nonumber \\ &-&\frac{2(L-M)(l-m)}{9\Delta}I_1 -\frac{4Nn}{3\Delta}I_2
    \end{eqnarray}
where
\begin{eqnarray} \label{c2}
    I_1&=&p_z^2(2u_{zz}-u_{xx}-u_{yy}) +p_x^2(2u_{xx}-u_{zz}-u_{yy}) \nonumber \\ &+&p_y^2(2u_{yy}-u_{xx}-u_{zz}),  \\ I_2&=&p_xp_yu_{xy}+p_zp_xu_{xz}+p_zp_yu_{yz}.   \label{c3}
    \end{eqnarray}
It is seen that the uniform compression leads to $I_1=I_2=0$, and we arrive at the formula obtained within the rigid-band approximation. The quantities $I_1$ and $I_2$ define the effect of the elastic deformation $u_{kl}$ on the ${\bf p}$ dependence of $\varepsilon_{h3}$. Note that the terms $I_1$ and $I_2$ are the only invariants of the point group $O_h$ that are quadratic in $p_i$ and linear in $u_{kl}$ and that vanish at the uniform compression.
In this context, the last two terms in Eq.~(\ref{c1}) could be immediately written as $A_1I_1+A_2I_2$ (with some constant coefficients $A_1$ and $A_2$) from symmetry considerations without a  reference to the formulas of the three-band model. However this model permits us to find the dependence of these coefficients on the gap $\Delta$.

In analyzing the magnetostriction, we may omit the term proportional to $I_2$ in Eq.~(\ref{c1}) and in all subsequent formulas (see the end of Appendix \ref{A}). The term proportional to $I_1$ in Eq.~(\ref{c1}) means that the deformation changes the mass $m_{h3}$, and in the deformed crystal, $\varepsilon_{h3}({\bf p})$ takes the form:
 \begin{eqnarray}\label{c4}
\varepsilon_{h3}({\bf p})=\frac{p_x^2}{2m_{h3}^{xx}}+\frac{p_y^2}{2m_{h3}^{yy}}+ \frac{p_z^2}{2m_{h3}^{zz}},
 \end{eqnarray}
with
\begin{eqnarray}\label{c5}
m_{h3}^{zz}&=& m_{h3}[1+2\kappa(2u_{zz}-u_{xx}-u_{yy})], \nonumber \\
m_{h3}^{xx}&=& m_{h3}[1+2\kappa(2u_{xx}-u_{zz}-u_{yy})], \\
m_{h3}^{yy}&=& m_{h3}[1+2\kappa(2u_{yy}-u_{xx}-u_{zz})], \nonumber \\
\kappa&\approx& \frac{2m_{h3}(L-M)(l-m)}{9\Delta}. \nonumber
\end{eqnarray}
Then, in the weak magnetic fields $B\ll F_{h3}$, we have the following expression for the contribution $\delta\Omega_{h3}=-\chi_{h3} B^2/2$ of the $h3$ holes to the $\Omega$ potential (we do not consider the oscillation part of $\delta\Omega_{h3}$, and $\chi_{h3}$ is the smooth part of the magnetic susceptibility):
 \begin{eqnarray}\label{c6}
\delta\Omega_{h3}=-\!\left(\frac{e}{c}\right)^2\!\!\frac{\sqrt{2m_{h3}^{zz} (\varepsilon_F+\Delta)}\,B^2}{2\pi^2\hbar\, |m_{h3}^{*}|} \!\left(\!\delta_{h3}^2-\frac{1}{12}\!\right)\!,
 \end{eqnarray}
where  $m_{h3}^{*}<0$ is the cyclotron mass in the deformed crystal, $|m_{h3}^{*}|=(m_{h3}^{xx}m_{h3}^{yy})^{1/2}$. Note that since $\delta_{h3}=g_{h3}m_{h3}^{*}/4m$ where $g_{h3}$ is the \textit{g} factor for the band $h3$, the combination in Eq.~(\ref{c6}),
 \[
\frac{\sqrt{|m_{h3}^{zz}|}\delta_{h3}^2}{|m_{h3}^{*}|}= \frac{\sqrt{|m_{h3}^{zz}|}|m_{h3}^{*}|g_{h3}^2}{16m^2}= \frac{|m_{h3}|^{3/2}g_{h3}^2}{16m^2},
 \]
depends on the deformation only through the \textit{g} factor. Taking into account the high symmetry of the point $\Gamma$, this dependence for the hole charge carriers looks like,
  \[
  g(u_{kl})=g({u_{kl}=0})[1+\alpha(u_{xx}+u_{yy}+u_{zz})],
  \]
where $\alpha$ is a constant of the order of $1$, and the dependence of the \textit{g} factor leads only to a small correction to the isotropic part of the magnetostriction produced by the holes (see Appendix \ref{B}). For this reason, we neglect the dependence of the hole $g$ factor on the deformation and omit $\delta_{h3}^2$ in Eq.~({\ref{c6}) when calculating $\partial(\delta \Omega_{h3})/\partial u_{ll}$ below.

To find a contribution associated with the dependence of the hole masses on the deformation to $u_{ll}^{(h3)}$,
we solve the first three  equations in (\ref{2a}) with the derivatives of $\delta\Omega_{h3}$ calculated at $\varepsilon_F$ independent of $u_{ll}$,
 \begin{eqnarray}\label{c7}
\frac{\partial\delta\Omega_{h3}}{\partial u_{zz}}&=&-2\frac{\partial\delta\Omega_{h3}}{\partial u_{xx}}=-2\frac{\partial\delta\Omega_{h3}}{\partial u_{yy}} \nonumber \\
&=&\!\left(\frac{e}{c}\right)^{5/2}\!
\frac{2\kappa\sqrt{2F_{h3}}\,B^2}{12\pi^2\hbar^{1/2}\, |m_{h3}|}\!,
 \end{eqnarray}
where we have also used relation (\ref{17b}).
Then,
 \begin{eqnarray}\label{c8}
u_{zz}^{(h3)}=-\frac{1}{(C_{33}-C_{13})}\frac{\partial\delta\Omega_{h3}}{\partial u_{zz}},\ \ \ \ u_{xx}^{(h3)}=-\frac{1}{2}u_{zz}^{(h3)}.~~~
 \end{eqnarray}
In the case of  arbitrary $B$ (including the case of $B>F_{h3}$), formula (\ref{c6}) must be replaced by a general expression for $\delta\Omega_{h3}$. Then, the last line in Eq.~(\ref{c7}) takes the form:
\begin{eqnarray} \label{c9}
 \frac{\partial\delta\Omega_{h3}}{\partial u_{zz}}\!&=&\!
 	-\frac{(e/c)^{5/2}2\kappa\sqrt{2} B^{5/2}}{\pi^2\hbar^{1/2}|m_{h3}|}\!\!\sum_{\pm}\!{\rm Im}\!\Biggl \{\!\frac 56 \zeta(-\frac 32,\, -u_{\pm} + \frac 12 ) \nonumber \\
  & +& \frac 12 \zeta(-\frac 12,\, -u_{\pm} + \frac 12 )u_{\pm}  \Biggr \},
\end{eqnarray}
where the notations of $u_{\pm}$ an $\zeta(-n,u)$ are the same as in Eq.~(\ref{23b}).

Consider now the bands $h1$ and $h2$ defined by Eq.~(\ref{15b}). Taking into account the above discussion of the band $h3$, we can write at once,
 \begin{eqnarray}\label{c10}
\varepsilon_{h1,h2}({\bf p})&=&\frac{p^2}{2m_{h1,h2}} +\lambda_h(u_{xx}+u_{yy}+u_{zz}) \nonumber \\
&+&\frac{(L-M)(l-m)}{9\Delta}I_1,
\end{eqnarray}
where the third term describes the effect of the $h3$ band on the bands $h1$ and $h2$. This term follows from the three-band model. Now formulas (\ref{c5}) take the form:
\begin{eqnarray}\label{c11}
m_{h1,h2}^{zz}&=& m_{h3}[1-\kappa_{h1,h2}(2u_{zz}-u_{xx}-u_{yy})], \nonumber \\
m_{h1,h2}^{xx}&=& m_{h3}[1-\kappa_{h1,h2}(2u_{xx}-u_{zz}-u_{yy})],~~~ \\
m_{h1,h2}^{yy}&=& m_{h3}[1-\kappa_{h1,h2}(2u_{yy}-u_{xx}-u_{zz})], \nonumber
\\
\kappa_{h1,h2}&=&\frac{m_{h1,h2}}{m_{h3}}\kappa.
\nonumber
\end{eqnarray}
Since the frequencies $F_{h1}$ and $F_{h2}$ are large, it is sufficient to calculate the magnetostriction produced by these pockets only in the range of the weak magnetic fields, $B\ll F_{h1}$, $F_{h2}$. Then, in formulas (\ref{c6}) and (\ref{c8}), the subscript (superscript) $h3$ is simply replaced by $h1$ or $h2$, and in the last line of formula (\ref{c7}), apart from this replacement, one should insert $-\kappa_{h1}$ or $-\kappa_{h2}$ instead of $2\kappa$.

The $u_{ll}^{(hi)}$ ($l=x,y,z$) derived in this Appendix should be added to the same quantities calculated within the rigid-band approximation in Appendix \ref{B}. It is clear that these total contributions of the holes to the magnetostriction are no longer isotropic, and hence the holes can contribute to $u_{zz}-u_{xx}$ measured in experiments. However, the $u_{ll}^{(hi)}$ of this Appendix have no effect on the isotropic part of the magnetostriction $(u_{zz}+2u_{xx})/3$ since the last of Eqs.~(\ref{c8}) shows that
$u_{zz}^{(hi)}+2u_{xx}^{(hi)}=0$. This cancelation of $u_{ll}^{(hi)}$ results from the symmetry of the crystal. Indeed, the $\Omega$ potential of the holes depends on $|m_{hi}^{zz}|$ and $|m_{hi}^{*}|=\sqrt{m_{hi}^{xx}m_{hi}^{yy}}$. Dependences of these masses on the deformation are determined by the unique combination $2u_{zz}-u_{xx}-u_{yy}$, see Eqs.~(\ref{c5}), (\ref{c11}), which leads to the relations in the first line of Eqs.~(\ref{c7}) and formulas (\ref{c8}). On the other hand, this combination is dictated by the cubic symmetry of the point $\Gamma$.

A similar analysis beyond the rigid-band approximation can be also  carried out for the electron pockets. This analysis introduces additional unknown constants of the deformation potential into the theory, but it does not lead to a qualitative change of the electron contribution to the magnetostriction, which is anisotropic even within the rigid-band approximation. For this reason, we do not consider the electron pockets beyond this approximation.

\section{Change of the quantum-oscillation frequencies under deformations} \label{D}

Values of the constants $\lambda_{zz}^{(c)}$, $\lambda_{zz}^{(a)}$, $\lambda_h$, and $\kappa$  can be obtained from independent experiments, measuring changes of frequencies $F_i$ under a uniform and uniaxial compressions of LuAs.

\subsection{Uniform compression}

In the case of uniform compression,  we have
\begin{eqnarray}\label{1d}
u_{xx}=u_{yy}=u_{zz}=-\frac{p}{C_{33}+2C_{13}},
\end{eqnarray}
where $p$ is the external pressure, and we find
the following formula for the derivative $dF_{h1,h2}/dp$:
\begin{eqnarray}\label{2d}
\frac{dF_{h1,h2}}{dp}=3\frac{cm_{h1,h2}}{e\hbar} \frac{\lambda_{zz}^{(h)}}{C_{33} +2C_{13}}=-3\frac{cm_{h1,h2}}{e\hbar}\Lambda_h,~~~
\end{eqnarray}
where $\Lambda_h$ is given by Eq.~(\ref{20b}), and $m_{h1,h2}$ are the hole masses (Appendix \ref{B}). This formula remains true even if we go beyond the rigid-band approximation, and it is not based on the simple model used for the holes in Appendices \ref{B} and \ref{C}. Equation (\ref{2d}) permits  one to find $\lambda_h$.

For the electron frequencies, the formula in the case of the uniform compression looks as follows:
\begin{eqnarray}\label{3d}
\frac{dF_i}{dp}=\frac{cm_i^*}{e\hbar}\frac{\lambda_{zz}^{(c)}+ 2\lambda_{zz}^{(a)}}{C_{33}+2C_{13}},
\end{eqnarray}
where $m_i^*$ are the appropriate cyclotron masses, and hence either $F_i=F_{c}$, $m_i^*=m_a$ or $F_i=F_{a}$, $m_i^*=(m_am_c)^{1/2}$; see Eqs.~(\ref{9b}). Note that formula (\ref{3d}) has been derived  in the rigid-band approximation. If this approximation fails for the electrons, the cyclotron masses $m_i^*$ will depend on $p$. In this situation, to obtain the correct formula, one should replace $d(F_i)/dp$  in the left hand side of Eq.~(\ref{3d}) by  $d(F_i/m_i^*)/dp$ and omit $m_i^*$ in its right hand side.

\subsection{Uniaxial compression}

For definiteness, let the uniaxial compression be applied to the sample along the $z$ axis. Then, the strain tensor has the form \cite{LL-sl}:
\begin{eqnarray}\label{4d}
 u_{zz}&=&-p\frac{C_{33}+C_{13}}{(C_{33}-C_{13})(C_{33}+2C_{13})}, \nonumber \\
u_{xx}&=&u_{yy}=p\frac{C_{13}}{(C_{33}-C_{13})(C_{33}+2C_{13})},
 \end{eqnarray}
where $p$ is the uniaxial pressure.

In the rigid-band approximation, we obtain the following formulas for the electron frequencies $F_c$ and $F_a$ when the magnetic field along $z$ (along the compression):
\begin{eqnarray}\label{5d}
\frac{dF_c}{dp}&=&\frac{cm_c^*}{e\hbar}\frac{[\lambda_{zz}^{(c)} (C_{33}+C_{13})- 2\lambda_{zz}^{(a)}C_{13}]}{(C_{33}-C_{13})(C_{33}+2C_{13})}, \\
\frac{dF_a}{dp}&=&\frac{cm_a^*}{e\hbar}\frac{[\lambda_{zz}^{(a)} C_{33}- \lambda_{zz}^{(c)}C_{13}]}{(C_{33}-C_{13})(C_{33}+2C_{13})}, \label{6d}
\end{eqnarray}
where $m_c^*=m_a$, and $m_a^*=(m_cm_a)^{1/2}$. If the magnetic field is along the $x$ axis ($B$ is perpendicular to the compression), we have
\begin{eqnarray}\label{7d}
\frac{dF_c}{dp}&=&\frac{cm_c^*}{e\hbar}\frac{[\lambda_{zz}^{(a)}C_{33}- \lambda_{zz}^{(c)} C_{13}]}{(C_{33}-C_{13})(C_{33}+2C_{13})}, \\
\frac{dF_a}{dp}&=&\frac{cm_a^*}{e\hbar}\frac{[\lambda_{zz}^{(c)}(C_{33} +C_{13}) -2\lambda_{zz}^{(a)}C_{13}]}{(C_{33}- C_{13})(C_{33}+2C_{13})}. \label{8d}
\end{eqnarray}
The visible symmetry of Eqs.~(\ref{5d}), (\ref{6d}) and (\ref{7d}),  (\ref{8d}) can be easily understood since the pocket $c$ becomes pocket $a$ and vice versa when the direction of $B$ changes from $z$ to $x$.
Knowing $dF_c/dp$ and $dF_a/dp$ for a certain direction of the magnetic field, one can find $\lambda_{zz}^{(c)}$ and $\lambda_{zz}^{(a)}$ for the electron pockets either from Eqs.~(\ref{5d}) and (\ref{6d}) or from Eqs.~(\ref{7d}) and (\ref{8d}).

Consider now the holes $h1$ and $h2$. If the the magnetic field is applied along the $z$ axis (along the uniaxial compression), we obtain
\begin{eqnarray}\label{9d}
\frac{dF_{h1,h2}}{dp}=\frac{cm_{h1,h2}}{e\hbar}\frac{\lambda_h} {(C_{33}+2C_{13})} - \frac{\kappa_{h1,h2}F_{h1,h2}} {(C_{33}-C_{13})},~~~~~~
\end{eqnarray}
whereas if $B$ is along the $x$, we arrive at
\begin{eqnarray}\label{10d}
\frac{dF_{h1,h2}}{dp}=\frac{cm_{h1,h2}}{e\hbar}\frac{\lambda_h} {(C_{33}+2C_{13})} + \frac{\kappa_{h1,h2}F_{h1,h2}} {2(C_{33}-C_{13})}.~~~~~~
\end{eqnarray}
Four equations (\ref{9d}) and (\ref{10d}) enable one to find $\kappa_{h1}$, $\kappa_{h2}$, and $\lambda_h$.


\begin{thebibliography}{}
\bibitem{c1} These considerations are not applicable to Weyl nodes that are  topologically protected. A small deformation cannot create a gap at such a node, and so one may expect the applicability of the rigid-band approximation to the case of the Weyl pockets \cite{cichorek}.
\end{thebibliography}
\end{document}